\documentclass[prd,aps,twocolumn,a4paper,showkeys,nofootinbib]{revtex4-1}

\usepackage{graphicx,psfrag}
\usepackage{mathrsfs}
\usepackage{amsmath,amsfonts,amssymb}
\usepackage{multirow}
\usepackage{comment}
\usepackage{subfigure}
\usepackage{appendix}
\usepackage{hyperref}
\usepackage{ragged2e}

\usepackage{enumitem}
\usepackage[normalem]{ulem}

\newcommand{\be}{\begin{equation}}
\newcommand{\ee}{\end{equation}}
\newcommand{\bea}{\begin{eqnarray}}
\newcommand{\eea}{\end{eqnarray}}
\newcommand{\bel}{\begin{align}}
\newcommand{\eel}{\end{align}}

\def\l{\ell}
\def\lm{\ell m}

\def\i{{\rm i}}

\def\GMc2{{\rm G M_{\odot} c^{-2}}}

\def\kt2{\kappa^\text{T}_2}

\usepackage{pifont} % \cmark \xmark
\newcommand{\cmark}{\ding{51}}%
\newcommand{\xmark}{\ding{55}}%

\newcommand{\BAM}[1]{\texttt{BAM:{#1}}} 

\newcommand{\EFL}[1]{EFL{#1}} 
\usepackage{color}
\definecolor{cyan}{rgb}{0,0.9,0.9}
\definecolor{orange}{rgb}{0.9,0.5,0}
\definecolor{magenta}{rgb}{1,0,1}
\definecolor{purple}{rgb}{0.8,0.4,0.8}
\definecolor{gray}{rgb}{0.8242,0.8242,0.8242}
 
\newcommand{\red}[1]{\textcolor{red}{#1}} 
\newcommand{\green}[1]{\textcolor{green}{#1}} 
\newcommand{\blue}[1]{\textcolor{blue}{#1}}
\newcommand{\purple}[1]{\textcolor{purple}{#1}}

\begin{document}

\title{Entropy-limited higher-order central scheme for neutron star merger simulations}

\author{Georgios \surname{Doulis}$^{1,2}$}
\author{Florian \surname{Atteneder}$^{1}$}
\author{Sebastiano \surname{Bernuzzi}$^{1}$}
\author{Bernd \surname{Br\"ugmann}$^{1}$}

\affiliation{${}^1$Theoretisch-Physikalisches Institut, Friedrich-Schiller-Universit{\"a}t Jena, 07743, Jena, Germany}
\affiliation{${}^2$Department of Physics, University of Athens, 15783, Athens, Greece}

\date{\today}

\begin{abstract}

Numerical relativity simulations are the only way to calculate 
exact gravitational waveforms from binary neutron star mergers 
and to design templates for gravitational-wave astronomy. The 
accuracy of these numerical calculations is critical in quantifying 
tidal effects near merger that are currently one of the main 
sources of uncertainty in merger waveforms. In this work, we 
explore the use of an entropy-based flux-limiting scheme for 
high-order, convergent simulations of neutron star spacetimes. 
The scheme effectively tracks the stellar surface and physical 
shocks using the residual of the entropy equation thus allowing 
the use of unlimited central flux schemes in regions of smooth 
flow. We perform the first neutron star merger simulations with 
such a method and demonstrate up to fourth-order convergence in 
the gravitational waveform phase. The scheme reduces the phase 
error up to a factor five when compared to state-of-the-art high-order 
characteristic schemes and can be employed for producing faithful 
tidal waveforms for gravitational-wave modelling.

\end{abstract}


\maketitle

\section{Introduction}

The detection of gravitational waves (GWs) from binary neutron star
(BNS) merger events by the LIGO-Virgo collaboration opened the way to
observationally probe NS matter with GW signals
\cite{TheLIGOScientific:2017qsa,Abbott:2018exr,Abbott:2018wiz,Abbott:2020uma}. 
Key to this endeavour is the availability of merger waveforms from
numerical relativity (NR) simulations that accurately resolve tidal
effects and allows the design of sophisticated waveform templates
\cite{Damour:2012yf,Baiotti:2010xh,Bernuzzi:2012ci,Read:2013zra,Bernuzzi:2015rla,Hinderer:2016eia,Dietrich:2017aum}. Current
tidal waveform templates have been shown to be inaccurate (unfaithful)
for the inference of tidal parameters at the signal-to-noise ratios
that would otherwise allow a precision measurement
\cite{Gamba:2020wgg}. The main source of inaccuracy is the modelling of
tidal interactions toward merger and it is 
related to the lack of sufficiently accurate NR simulations. This is a
critical open issue for science with advanced detectors and an urgent
problem to solve in view of third-generation \cite{Kalogera:2021bya}.

Current state-of-the-art\footnote{%
  We focus here on the numerical quality and do not discuss other
  important aspects like eccentricity reduced circular initial data
  \cite{Kyutoku:2014yba,Dietrich:2015pxa}, the exploration of
  mass ratio \cite{Dietrich:2016hky,Bernuzzi:2020txg}, spin effects
  \cite{Bernuzzi:2013rza,Dietrich:2017xqb}
  generic orbits \cite{Gold:2011df}, or the influence of microphysics
  \cite{Radice:2018pdn,Nedora:2020pak}.
}
NR waveforms for modeling tidal interactions span about ten orbits to
merger and have typical accumulated phase errors below one radiant 
e.g.~\cite{Bernuzzi:2012ci,Radice:2013hxh,Hotokezaka:2015xka,Bernuzzi:2016pie,Dietrich:2018upm}.
Early studies pointed to numerical dissipation in relativistic
hydrodynamics (GRHD), to the numerical handling of the
stellar surfaces and to the slow convergence of high-resolution
shock-capturing (HRSC) as the main difficulties towards the computation of precise waveforms~\cite{Baiotti:2009gk,Thierfelder:2011yi,Bernuzzi:2011aq,Bernuzzi:2012ci,Hotokezaka:2013mm}.
The primary goal is to assess waveforms' error budget based on
convergent data and rigorous self-convergence tests, that has been
presented by few groups~\cite{Thierfelder:2011yi,Bernuzzi:2011aq,Radice:2013hxh}.
Traditional finite volume methods for GRHD using linear
reconstructions, the piece-wise parabolic method
\cite{Colella:1982ee,Marti:1996} or even third-order 
convex-essentially-non-oscillatory (CENO3) algorithm 
\cite{Liu:1998,Zanna:2002} allows robust and successful 
simulations but do not produce convergent waveforms at 
affordable resolutions \cite{Giacomazzo:2009mp,Thierfelder:2011yi,Bernuzzi:2012ci,Radice:2013xpa}.
Consequently, high-order (HO) numerical schemes based on 
fifth-order characteristic reconstructions of the GRHD fields 
\cite{Shu:1988} have been explored and represent the best 
methods available to date \cite{Radice:2013hxh,Bernuzzi:2016pie}.
HO schemes allow the computation of convergent waveforms but none of
the schemes tested so far achieves the formal high-order accuracy
expected for smooth flow. Nonetheless, the direct data comparison
between two independent codes indicate good agreement within the
estimated errorbars \cite{Nagar:2018zoe} (see Appendix~D). 
While convergent waveform can be obtained, the computational cost of
producing GW with sub-radiant accuracy over multiple orbits and to
merger remains rather high \cite{Radice:2015nva,Dietrich:2018upm}. 

In the present work we explore further the potential  
of a method that started as an artificial viscosity 
method \cite{Guermond:2008} and developed through the 
years to a flux-limiting method \cite{Guercilena:2016fdl}. 
The central idea of this method is to use a physical 
quantity as an indicator of the location of abnormal 
non-smooth regions like shocks, rarefactions etc. 
Entropy is an ideal candidate for this role as shocks 
are irreversible processes and thus increase the overall 
entropy of the system. Therefore, entropy can be used 
to flag the presence of non-smooth features in the 
solution space. The idea of using the entropy to design 
numerical methods for non-linear conservation laws 
is not new though. For example it is shown in 
\cite{Andrews:1998,Puppo:2004} that the entropy 
production can be used as a posteriori shock indicator 
and therefore it is extremely useful in the shock 
tracking. The authors in \cite{Guermond:2008,Guermond:2011} 
used the aforementioned idea to design a novel class 
of high-order numerical approximations to non-linear 
conservation laws by adding a degenerate non-linear 
dissipation to the numerically discretized system. 
The additional non-linear viscosity term is based on 
the local size of the entropy production. By making 
the numerical diffusion proportional to the 
entropy production in strong shocks, large numerical 
dissipation is added in the shock regions and almost 
no dissipation in the regions where the solution is 
smooth. This close interplay between the notions of 
entropy and viscosity gave the name entropy-viscosity 
(EV) to this method. 

In \cite{Guercilena:2016fdl} the EV method was 
incorporated in a HRSC method and extended 
to special and general relativistic hydrodynamics. 
Accordingly, the definitions of the entropy and viscosity 
were generalised and the viscosity was employed to 
drive a flux-limiting scheme rather than generating 
additional viscous terms in the hydrodynamical equations. 
The equations of GRHD are not modified anymore by 
the inclusion of additional viscosity related terms. 
Instead, a flux-limiting strategy is employed, i.e. 
the numerical fluxes are computed using an unlimited high-order stencil complemented by a first-order, 
non-oscillatory local Lax-Friedrichs (LLF) flux in 
regions of non-smooth flow. The high-order and low-order 
fluxes are linearly combined using local weights that 
are determined by i) an entropy-based shock detector 
criterion (based on the residual of the entropy equation) 
and ii) a positivity preserving limiter \cite{Hu:2013}. 
This hybrid scheme was named entropy-limited hydrodynamics 
(ELH). It has been shown effective in capturing shocks 
and discontinuities in special relativistic shock tubes 
as well as in producing stable evolutions of single 
neutron star by properly handling stellar surface effects. 
\cite{Guercilena:2016fdl} also points out shortcomings 
of the method: small spurious oscillations are observed 
in the blast wave 2 test, while neutron star evolutions 
show a spurious direction-dependent feature that breaks 
spherical symmetry.

In the present work, we build upon the existing 
machinery of the ELH method. While keeping loyal to 
the basic features of the method, we extend and 
generalise some of its aspects and modify or even 
drop some others. Most noticeably we drop the 
use of the positivity preserving limiter \cite{Hu:2013} 
and define the weights of the fluxes directly from 
the entropy produced by the system under investigation. 
Another notable amendment is that we allow the unfiltered 
high-order flux to be supplemented by general stable 
low- or high-order fluxes. In addition, the ELH is 
simplified by completely defining the free parameters 
inherent in the method. In light of the above quantitative 
differences with the ELH method we name the scheme 
developed in the present work \textit{entropy based flux-limiter} 
(EFL) method as it describes exactly what we have 
developed: a genuine entropy based flux-limiter. The 
new, EFL scheme remains robust in handling the special 
relativistic and the single neutron star tests, notably 
improving the shortcoming of the previous implementation. 
Moreover, we successfully apply for the first time the 
scheme to BNS simulations.We discuss high-order convergence 
in the inspiral-merger GWs and future prospect for 
producing faithful waveforms for GW modeling.

The article is structured as follows. In \autoref{sec:method}, 
after briefly summarising the equation of GRHD, we 
discuss theoretical and numerical aspects of our method. 
\autoref{sec:SRHD_tests} includes our results for the 
standard benchmark tests of special relativity, and in 
\autoref{sec:SNS} the performance of our method is 
tested against three-dimensional general relativistic 
single NS configurations. Our main results are presented 
in \autoref{sec:BNS}, where the first BNS evolutions with 
a method based on the entropy production can be found. 
Finally, we conclude in \autoref{sec:conclusions}.

Throughout this work we use geometric units. We set 
$c = G = 1$ and the masses are expressed in terms of 
solar masses $M_\odot$.

\section{Method}
\label{sec:method}

\subsection{General relativistic hydrodynamics}
\label{sec:grhd}

The evolution of a relativistic fluid in the presence of a 
non-trivial gravitational field $g_{\mu\nu}$ is described 
by the local conservation laws of the energy-momentum tensor 
$T_{\mu\nu}$ and of the rest-mass current $J^\mu = \rho\,u^\mu$,
\begin{equation}
 \label{eq:conserv_laws}
 \nabla_\mu T^{\mu\nu} = 0 \quad \mathrm{and} \quad \nabla_\mu J^\mu = 0,
\end{equation}
respectively. Above $\nabla$ denotes the covariant derivative 
compatible with $g_{\mu\nu}$, $\rho$ is the rest-mass density 
and $u^\mu$ is the 4-velocity of the fluid. The evolution 
equations \eqref{eq:conserv_laws} in the 3+1 formalism 
\cite{Alcubierre:2008} can be written as a system of PDEs 
in conservation form \cite{Banyuls:1997zz}
\begin{equation}
 \label{eq:conserv_PDE}
 \partial_t \textbf{Q} + \partial_i \textbf{F}^i = \textbf{S},
\end{equation}
where the summation is performed over the spatial dimensions 
$i=\{ x, y, z\}$ and the vector $\textbf{Q}$ of the conserved 
variables reads
\begin{align}
 \label{eq:conserv}
  \textbf{Q} = \sqrt{\gamma} \begin{bmatrix}
                              D \\
                              S_j \\
                              \tau
                             \end{bmatrix}
             := \sqrt{\gamma} \begin{bmatrix}
                               \rho\,W \\
                               \rho\,h\,W^2\,u_j \\
                               \rho\,h\,W^2-p-\rho\,W
                              \end{bmatrix}
\end{align}
where $S_j = \{ S_x,S_y,S_z\}$, $p$ is the pressure, $h$ 
is the specific enthalpy $h=1 + \epsilon + p/\rho$ with 
$\epsilon$ the specific internal energy, $W = (1 - u_i\,u^i)^{1/2}$ 
is the Lorentz factor and $\gamma$ is the determinant of 
the 3-metric $\gamma_{ij}$ resulting from the 3+1 decomposition 
of ($\mathcal{M}, g_{\mu\nu}$). The vector $\textbf{F}^i$ 
of the physical fluxes is
\begin{align}
 \label{eq:phys_flux}
  \textbf{F}^i  
  = \sqrt{\gamma} 
  \begin{bmatrix}
    (\alpha\,u^i-\beta^i)\,D \\
    (\alpha \,u^i-\beta^i)\, S_j + \alpha\,p\,\delta^i_j\\
    (\alpha \,u^i-\beta^i)\,\tau + \alpha\,p\,u^i
  \end{bmatrix},
\end{align}
where $\alpha$ is the lapse function, $\beta^i$ the shift 
vector and $\delta^i_j$ the Kronecker delta. Finally, the 
vector $\textbf{S}$ of the sources has the form 
\begin{align}
 \label{eq:sourse}
  \textbf{S} = \alpha\,\sqrt{\gamma} 
      \begin{bmatrix}
       0 \\
       \Gamma^\mu_{\nu j} T^\nu_\mu\\
       \alpha\left(T^{0\mu} \partial_\mu \ln\alpha-\Gamma^0_{\mu\nu} T^{\mu\nu}\right)
      \end{bmatrix},
\end{align}
where $\Gamma^\rho_{\mu\nu}$ are the Christoffel symbols 
associated with the metric $g_{\mu\nu}$. Notice that the 
system \eqref{eq:conserv_PDE} reduces to its special 
relativistic counterpart in the limit $(\alpha,\beta^i,
\gamma_{ij}) \rightarrow (1,0,\delta_{ij})$, i.e.\ when 
$\textbf{S} \rightarrow 0$. 

In order to close the underdetermined system \eqref{eq:conserv_PDE} 
one needs an equation of state (EoS) that specifies the 
pressure in terms of the density and the internal energy, 
i.e $p=p(\rho,\epsilon)$. Specifically, for the special 
relativistic tests of Sec.~\ref{sec:SRHD_tests} we use a 
$\Gamma$-law EoS, 
\begin{equation}
 \label{eq:gamma_law}
 p = (\Gamma - 1) \rho\, \epsilon,
\end{equation}
with $\Gamma$ the adiabatic index. The neutron star matter 
of the single neutron star evolutions of Sec.~\ref{sec:SNS} 
is also modelled by a $\Gamma$-law EoS \eqref{eq:gamma_law}. 
Finally, the matter of the neutron stars comprising the 
binaries of Sec.~\ref{sec:BNS} is described by either 
a $\Gamma$-law EoS \eqref{eq:gamma_law} or by a more 
realistic SLy EoS \cite{Douchin:2001sv}. The latter is 
implemented by a piecewise polytrope fit \cite{Read:2008iy}, 
and thermal effects are modeled by an additive pressure 
contribution given by the $\Gamma$-law EoS with $\Gamma 
= 1.75$ \cite{Shibata:2005ss,Bauswein:2010dn,Thierfelder:2011yi}. 

\subsection{EFL method}
\label{sec:EFL_method}

In the present work the entropy-viscosity (EV) method 
\cite{Guermond:2008, Guermond:2011} is used as a flux 
limiting scheme in the spirit of \cite{Guercilena:2016fdl}. 
In \cite{Guercilena:2016fdl} the original EV method 
was reformulated as an entropy based flux-limiter and 
extended to special and general relativistic hydrodynamics. 
The basic idea of the ELH method consists of expressing 
the numerical fluxes resulting from the spatial discretisation 
of \eqref{eq:conserv_PDE} as a superposition of an 
(unstable) high- and a (stable) low-order flux, where 
the weight dictating the transition between the two 
fluxes is computed based on the entropy produced in 
the system under investigation. The entropy of the 
system is used as a ``shock detector" that indicates 
when to switch from the high-order scheme to the 
low-order one. 

The EFL method follows in broad lines the exposition 
in \cite{Guercilena:2016fdl}, but adds some novelties 
to the already existing scheme. One of the main differences 
is that we do not use the positivity-preserving limiter 
\cite{Radice:2013xpa} in the definition of the transition 
weight $\theta$. Another key development is that the LO 
flux here is composed of a non-oscillatory high-order 
scheme, namely a finite volume method with high-order 
reconstruction (CENO3, WENO, etc.). In this way the 
chances that the resulting hybrid flux can achieve 
high-order convergence rates are maximised. Finally, 
the handling of the tunable constants is extremely 
simplified, see last paragraph of the current section 
for further details.

We start by approximating the spatial derivative of 
the $x$ component, $\textbf{F}^x$, of the physical 
flux \eqref{eq:phys_flux} appearing in \eqref{eq:conserv_PDE} 
with the conservative finite-difference formula\footnote{For 
clarity and without loss of generality, from now on the 
presentation is restricted to one dimension, say $x$. 
A multidimensional scheme is obtained by considering 
fluxes in each direction separately and adding them 
to the r.h.s.}
\begin{equation}
 \label{eq:spatial_disc}
  \partial_x F^x_i = \frac{\hat f_{i+1/2} - \hat f_{i-1/2}}{h},
\end{equation}
where $F^x$ is any one of the components of $\textbf{F}^x$ 
with $F^x_i=F^x(x_i)$, $\hat f_{i\pm1/2}$ are the numerical 
fluxes at the cell interfaces and $h$ is the spatial 
grid spacing.

Next, we split the numerical fluxes on the r.h.s.\ of
\eqref{eq:spatial_disc} into two contributions, see also
\cite{Guercilena:2016fdl}: one from a HO scheme and one 
from a low-order (LO) stable scheme, i.e.
\begin{equation}
 \label{eq:num_flx_split}
  \hat f_{i\pm1/2} = \theta_{i\pm1/2} \hat f^{\,\mathrm{HO}}_{i\pm1/2} + 
  (1-\theta_{i\pm1/2}) \hat f^{\,\mathrm{LO}}_{i\pm1/2},
\end{equation}
where the continuous parameter $\theta \in [0,1]$ plays 
the role of a weight that indicates how much from each
scheme to use at every instance. The HO flux $\hat f^{\,\mathrm{HO}}$ 
is built using the Rusanov Lax-Friedrichs flux-splitting 
technique and performing the reconstruction on the 
characteristic fields \cite{Mignone:2010br,Bernuzzi:2016pie}. 
A fifth-order central unfiltered stencil (CS5) is always 
used for reconstruction. The LO flux $\hat f^{\,\mathrm{LO}}$ 
is approximated by the LLF central scheme with reconstruction 
performed on the primitive variables \cite{Thierfelder:2011yi}. 
Primitive reconstruction is performed with a variety 
of low- and high-order reconstruction schemes. (Notice 
that we generalise the traditional notion of a flux-limited 
scheme where $\hat f^{\,\mathrm{LO}}$ is always a LO 
monotone flux \cite{Toro:1999,Hesthaven:2017}.) A list 
of the ones used in the present work follows. Godunov's 
piecewise constant reconstruction scheme 
(GODUNOV) \cite{Godunov:1959}; the second-order linear 
total variation diminishing (LINTVD) interpolation 
based on ``minmod'' and ``monotonized centered'' slope 
limiters \cite{Harten:1982kq,Toro:1999}; the third-order 
convex-essentially-non-oscillatory (CENO3) algorithm 
\cite{Liu:1998,Zanna:2002}; and the fifth-order 
weighted-essentially-non-oscillatory finite difference 
schemes WENO5 \cite{Jiang:1996} and WENOZ \cite{Borges:2008a}. 
As it was mentioned above, this is a basic difference 
of the EFL method with the one proposed in \cite{Guercilena:2016fdl}; 
therein a first-order Lax-Friedrichs flux was used 
exclusively as the LO flux.
    
The computation of $\theta$ is based on the so-called 
\textit{entropy production function} $\nu$: a quantity 
that depends on the amount of entropy produced in the 
system. Explicitly, the relation between $\theta$ and 
$\nu$ is
\begin{equation}
 \label{eq:def_theta}
  \theta_{i\pm1/2} = 1- \frac{1}{2} (\nu_i + \nu_{i\pm1}).
\end{equation}   
Below, we summarise how to compute $\nu$. 

In order to quantify the relation between $\nu$ and 
the entropy produced by the system under investigation, 
we define the specific entropy (entropy per unit mass) 
of any piecewise polytropic EoS\footnote{For a more 
general EoS the specific entropy $s$ can be taken from 
the EoS.} as
\begin{equation}
 \label{eq:spec_entrp}
  s = \ln \left( \frac{p}{\rho^\Gamma} \right),
\end{equation}
where the pressure is computed in accordance with the 
EoS in use.

Following~\cite{Guercilena:2016fdl}, we employ the second 
law of thermodynamics to define the entropy residual: 
\begin{equation}
  \mathcal{R} = \nabla_\mu (s\,\rho\,u^\mu) \geq 0,
\end{equation}
which provides a quantitative estimation of the rate 
of the entropy produced by the system under study. Using 
the continuity equation and writing the 4-velocity $u^\mu$ 
in terms of the fluid 3-velocity $\upsilon^i$, the above 
expression can be written \cite{Guercilena:2016fdl} in 
terms of the time and spatial derivatives of the specific 
entropy as
\begin{equation}
 \label{eq:phys_entrp_resid_GRR}
 \mathcal{R} = \frac{\rho W}{\alpha}
 \left(\partial_t s + (\alpha\, \upsilon^i - \beta^i) \partial_i s \right).
\end{equation}
In order to simplify the definition of the constant 
$c_E$, see discussion below, we suppress the multiplication 
factor $\frac{\rho W}{\alpha}$ and replace $\mathcal{R}$ 
by
\begin{equation}
  \label{eq:phys_entrp_resid}
  R = \partial_t s + (\alpha\, \upsilon^i - \beta^i) \partial_i s,
\end{equation}
which amounts to a rescaling of $\mathcal{R}$ so that
the coefficient of $\partial_t s$ is equal to one.

Finally, we define the \textit{entropy production function} 
in terms of the rescaled entropy residual $R$,
\begin{equation}
 \label{eq:nu_E}
  \nu_E = c_E |R|,
\end{equation}
where $c_E$ is a tunable constant used to scale the absolute 
value of $R$. In all our simulations we 
did not have to tune $c_E$, its value was set to unity, 
i.e.\ $c_E=1$. Keeping in mind that the parameter $\theta$ 
cannot exceed unity, we have to impose a maximum value of 
$\nu_{\rm max}=1$ for the entropy production function in order 
to ensure that the rhs of \eqref{eq:def_theta} does not 
exceed the range $[0,1]$. Accordingly, the entropy production 
function entering \eqref{eq:def_theta} is given by 
\begin{equation}
 \label{eq:nu}
  \nu = \min\left[\nu_E, 1\right].
\end{equation}

Comparing directly with \cite{Guercilena:2016fdl}, note the following
differences.  In the present work, we use \eqref{eq:def_theta}
directly for the definition of $\theta$, while
\cite{Guercilena:2016fdl} adds a condition for positivity
preservation. We define $R$ as in \eqref{eq:phys_entrp_resid}, while
\cite{Guercilena:2016fdl} considers $\mathcal{R}=\frac{\rho W}{\alpha}
R$. Finally, we define the entropy production function as $\nu_E =
c_E |R|$, while in \cite{Guercilena:2016fdl} $\nu_E$ is multiplied
with $\Delta_m$, where $\Delta_m$ is the mesh spacing. 

In other words, based on various numerical experiments we found it
advantageous to remove the factor $\frac{\rho\, W \Delta_m}{\alpha}$
from the definition of the entropy production function $\nu_E$ compared 
to ELH. We study results for $c_E=1$ in detail, while
$c_E=\frac{\rho\, W\Delta_m}{\alpha}$ is considered in \cite{Guercilena:2016fdl}.
In the EFL method proposed here, there is no direct resolution
dependence, and the entropy production has been normalized to the
scale of $\partial_t s$.

\subsection{Numerical implementation}
\label{sec:num_implement}

The finite differencing code BAM \cite{Brugmann:2008zz,
Thierfelder:2011yi,Dietrich:2015iva,Bernuzzi:2016pie} 
is used to solve numerically the system of equations 
discussed in \autoref{sec:grhd} coupled to the metric 
equations for General Relativity. The EFL method presented 
in \autoref{sec:EFL_method} has been implemented into 
BAM and is part of its infrastructure. BAM uses the 
method-of-lines with Runge-Kutta (RK) time integration and 
finite differences for the approximation of spatial 
derivatives. The value of the Courant-Friedrich-Lewy (CFL) 
condition is set to 0.25 for all runs.

The numerical domain contains a mesh made of a hierarchy 
of cell-centered nested Cartesian boxes and consists 
of $L$ refinement levels $l = 0,\dots,L-1$ ordered 
with increasing resolution. Each refinement level is 
made out of one or more equally spaced Cartesian grids 
with grid spacing $h_l$. There are $n$ points per 
direction on each grid plus a certain number of buffer 
points on each side. (For simplicity, we always quote 
grid sizes without buffer points.) The resolution between 
two consecutive levels is doubled such that the grid 
spacing at level $l$ is $h_l = h_0/2^l$, where $h_0$ 
is the grid spacing of the coarsest level. The inner 
levels move in accordance with the moving boxes technique, 
while the outer levels remain fixed. The number of 
points in one direction of a moving level can be set 
to a different value than the number of points of a 
fixed level. The coordinate extent of a grid at level 
$l$ entirely contains grids at any level greater than 
$l$. The moving refinement levels always stay within 
the coarsest level. For the time evolution of the grid 
the Berger-Oliger algorithm is employed enforcing mass 
conservation across refinement boundaries 
\cite{Dietrich:2015iva,Berger:1984zza}. Restriction 
and prolongation is performed for the matter fields 
with a fourth-order WENO scheme and for the metric 
fields with a sixth-order Lagrangian scheme. Interpolation 
in Berger-Oliger time stepping is performed at second-order. 

For the numerical implementation of the EFL method the 
BAM routines computing the numerical fluxes had to be 
modified in order to accommodate the hybrid flux 
\eqref{eq:num_flx_split}. In order to compute the 
entropy production \eqref{eq:phys_entrp_resid} we have 
to approximate the time and spatial derivatives 
of the specific entropy. We use finite differences to do 
so. Specifically, the spatial derivatives are approximated, 
as in \cite{Guercilena:2016fdl}, with a standard centered 
finite-difference stencil of order $p+1$ or higher, where 
$p$ is the order of the stencil used to approximate the 
physical fluxes. (In the present work we use $p=5$.) 
With this restriction it is ensured that the entropy 
production function $\nu$ converges to zero faster than 
the overall convergence of the scheme. The time derivative 
is also approximated with finite differences. We employ 
a third-order one-sided stencil by using, at every 
Runge-Kutta iteration, the current value of the specific 
entropy and the values at the three previous timesteps. 
The fact that we manage to achieve higher than third-order 
convergence in the majority of our simulations can be 
possibly attributed to the dominance of the spatial 
error over the time discretization error. 

The derivatives of the metric components are approximated 
by fourth-order accurate finite-differencing stencils. 
In addition, sixth-order artificial dissipation operators 
are employed to stabilize noise from mesh refinement 
boundaries. The general relativistic hydrodynamic 
equations \eqref{eq:conserv_PDE} are solved by means 
of a high-resolution-shock-capturing method \cite{Thierfelder:2011yi} 
based on primitive reconstruction and the aforedescribed 
high-order entropy limited scheme for the numerical 
fluxes. In the present work spacetime is dynamically 
evolved using either the BSSNOK 
\cite{Nakamura:1987zz,Shibata:1995we,Baumgarte:1998te} 
or the Z4c \cite{Bernuzzi:2009ex,Hilditch:2012fp} evolution 
scheme.

Vacuum regions are simulated with the introduction of 
a static, low-density, cold atmosphere in the vacuum 
region surrounding the star \cite{Thierfelder:2011yi}. 
The atmosphere density is defined as 
\begin{equation}
 \label{eq:atmos}
 \rho_{\rm atm} = f_{\rm atm}\, \mathrm{max}\, \rho(t = 0).
\end{equation}
All grid points with rest-mass density below a threshold 
value $\rho_{\rm thr} = f_{\rm thr}\rho_{\rm atm}$ are set 
automatically to $\rho_{\rm atm}$. Transition to low-density 
regions is one of the main sources of error in NS simulations. 
This is a common feature in all current numerical relativity 
implementations of NS dynamics. To deal with this challenging 
feature they also make use of similar assumptions and 
algorithms at low densities as those employed here. We leave 
it to future work to investigate whether the advantages of
the atmosphere and vacuum treatment of \cite{Poudel:2020},
which improved mass conservation and accuracy of ejecta in 
that case study, can be combined with the new flux-limiting 
scheme. In the present work, we use the standard atmosphere 
treatment implemented in BAM \cite{Thierfelder:2011yi}, as 
our aim is to compare the performance of the newly developed 
entropy based flux-limiting scheme with our current best 
high-order flux scheme \cite{Bernuzzi:2016pie}.

\begin{figure}[t]
 \centering 
  \includegraphics[width=0.45\textwidth]{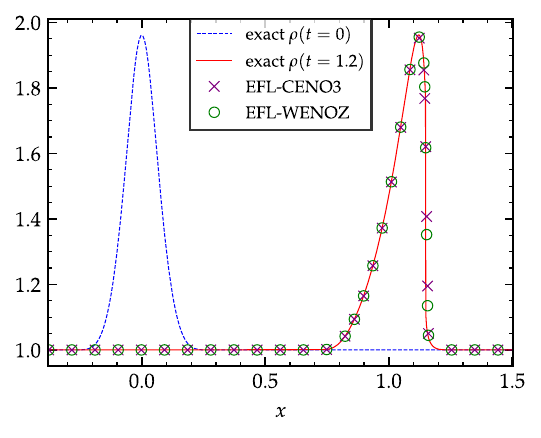}
   \caption{Simple wave. Numerical solution of the rest-mass 
   density with $n=800$ grid-points and CFL factor 0.125 for 
   the reconstruction schemes CENO3 (using purple \purple{$\times$})
   and WENOZ (using green \green{$\circ$}). The initial profile 
   and the exact solution are also included using dashed blue 
   and solid red lines, respectively.}
 \label{fig:simple_wave}
\end{figure}

\section{Special relativistic 1D tests}
\label{sec:SRHD_tests}

In this section a number of special relativistic one-dimensional 
tests are performed.

\subsection{Simple wave}

\begin{table}[t]
 \centering    
 \caption{Convergence results for the 1D simple wave test 
 at $t = 0.6$. $L_1$ and $L_2$ are normalised norms and 
 the convergence rate is calculated as the $\log_2$ of the 
 ratio of two successive normalised norms.}
   \begin{tabular}{cccccc}        
    \hline
    \hline
    Scheme    & n   & $L_1$   & Conv.& $L_2$   & Conv.\\
    \hline
    \EFL{-WENO5}     & 200 & 5.8e-04 & --   & 1.9e-04 &  --  \\
              & 400 & 2.6e-05 & 4.47 & 7.6e-06 & 4.66 \\
              & 800 & 1.2e-06 & 4.41 & 2.6e-07 & 4.87 \\
              & 1600& 6.4e-08 & 4.26 & 6.9e-09 & 5.22 \\
              & 3200& 7.7e-09 & 3.05 & 3.6e-10 & 4.28 \\
    \hline
   \EFL{-WENOZ}      & 200 & 5.5e-04 & --   & 1.8e-04 &  --  \\
              & 400 & 2.6e-05 & 4.42 & 7.5e-06 & 4.57 \\
              & 800 & 1.2e-06 & 4.39 & 2.6e-07 & 4.87 \\
              & 1600& 6.4e-08 & 4.26 & 6.9e-09 & 5.22 \\
              & 3200& 7.7e-09 & 3.05 & 3.6e-10 & 4.28 \\
    \hline
    \EFL{-CENO3}     & 200 & 6.9e-04 & --   & 2.3e-04 &  --  \\
              & 400 & 3.1e-05 & 4.50 & 8.5e-06 & 4.79 \\
              & 800 & 1.5e-06 & 4.37 & 2.6e-07 & 5.01 \\
              & 1600& 9.1e-08 & 4.01 & 7.7e-09 & 5.09 \\
              & 3200& 1.1e-08 & 3.10 & 4.8e-10 & 4.01 \\
    \hline
    \EFL{-LINTVD}    & 200 & 1.0e-03 & --   & 3.6e-04 &  --  \\
              & 400 & 3.4e-05 & 4.88 & 1.0e-05 & 5.14 \\
              & 800 & 1.4e-06 & 4.57 & 2.9e-07 & 5.14 \\
              & 1600& 1.0e-07 & 3.79 & 1.0e-08 & 4.84 \\
              & 3200& 1.3e-08 & 3.00 & 7.6e-10 & 3.76 \\
    \hline
    HO-WENOZ  & 200 & 4.4e-04 & --  & 1.3e-04 &  --   \\
              & 400 & 2.8e-05 & 3.98& 7.2e-06 & 4.16  \\
              & 800 & 1.2e-06 & 4.53& 2.6e-07 & 4.81  \\
              & 1600& 4.5e-08 & 4.73& 6.5e-09 & 5.30  \\
              & 3200& 5.7e-09 & 2.98& 2.8e-10 & 4.57  \\
    \hline
    \hline
   \end{tabular}
 \label{tab:simple_conv}
\end{table} 

The relativistic simple wave is used as a first check of 
the accuracy and of the convergence properties of the EFL 
method. Although, simple waves start off from smooth initial 
data, their non-linear nature leads to the development of 
shocks at some point during their evolution. These tests 
have been discussed in \cite{Liang:1977a,Anile:1990a}. 
Here, we use the simple wave described in \cite{Bernuzzi:2016pie}, 
therein the initial velocity profile is of the form
\begin{equation}
 \label{eq:v_profile}
  \upsilon = a\, \Theta(|x| - X) \sin^6 \left( \frac{\pi}{2} \left( 
  \frac{x}{X} -1 \right) \right),
\end{equation} 
where $\Theta(x)$ is the Heaviside function, $a = 0.5$ and 
$X = 0.3$. During the evolution the smooth initial profiles 
of all primitive variables become steeper and steeper and 
at around $t \simeq 0.63$ they form a shock. We use exactly 
the same numerical set-up with \cite{Bernuzzi:2016pie},
i.e.\ our one-dimensional computational domain spans the interval 
$x \in [-1.5, 1.5]$, RK4 is used as time-integrator and a 
CFL factor of 0.125 has been chosen. \autoref{fig:simple_wave} 
depicts the simple wave at $t = 1.2$ for a resolution of 
800 grid-points ($h = 0.00375$) for a high-order WENOZ 
and a lower-order CENO3 reconstruction scheme. (The behaviour 
of the other two reconstruction schemes used in this work 
is identical to the one depicted by \autoref{fig:simple_wave}.) 
By inspection, all schemes reproduce the correct physics. 
\autoref{tab:simple_conv} contains the results of the 
convergence analysis of the EFL schemes of 
\autoref{fig:simple_wave} at $t = 0.6$ (just before the 
shock forms). As a reference the HO-WENOZ scheme developed 
in \cite{Bernuzzi:2016pie} is also included in 
\autoref{tab:simple_conv}---this is the high-order 
scheme that we use to approximate the HO flux $\hat f^{\,\mathrm{HO}}$ 
in \eqref{eq:num_flx_split}, but with WENOZ (instead of 
CS5) for the reconstruction of the characteristic variables.
All schemes converge to the exact solution with the expected 
convergence rate.

\subsection{Sod shock-tube}

\begin{figure}[t]
 \centering 
  \includegraphics[width=0.45\textwidth]{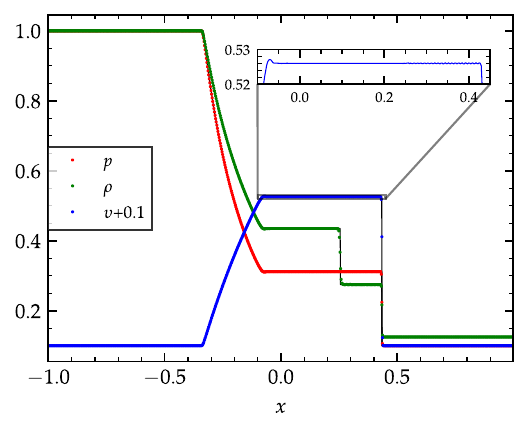}
  \put(-150,184){Sod: \EFL{-WENO5}}\\
  \vspace{-2mm}
  \includegraphics[width=0.45\textwidth]{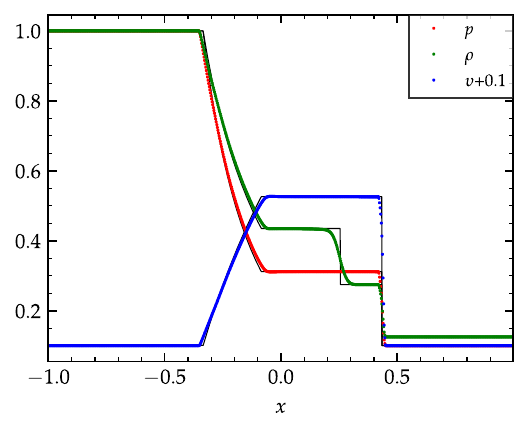}
  \put(-155,184){Sod: \EFL{-GODUNOV}}
   \caption{Profiles of the rest-mass density (green \green{\textbullet}), 
   velocity (blue \blue{\textbullet}) and pressure (red \red{\textbullet}) 
   for the special-relativistic Sod test at $t = 0.6$. Top: WENO5 
   reconstruction. Bottom: GODUNOV reconstruction. The solution 
   is computed on a grid of 1600 points with resolution $\Delta x 
   = 1.25 \times 10^{-3}$. Solid black lines are the exact solutions.}
 \label{fig:Sod}
\end{figure}  

We move on now to the standard Riemann problems used 
as benchmarks in special relativistic hydrodynamics. 
Our first test is the relativistic version of Sod's 
shock-tube problem \cite{Sod:1978}. Assuming a simple 
ideal fluid EoS of the form \eqref{eq:gamma_law} with 
adiabatic index $\Gamma = 1.4$, the discontinuous initial 
data for the pressure $p$, the rest-mass density $\rho$, 
the velocity $\upsilon$, and the specific energy $\epsilon$ 
read
\begin{equation}
 \label{eq:Sod_id}
  \begin{aligned}
   ( p_L, \rho_L, \upsilon_L, \epsilon_L ) &= (1,   1,     0, 2.5 ), \\ 
   ( p_R, \rho_R, \upsilon_R, \epsilon_R ) &= (0.1, 0.125, 0, 2 ).
  \end{aligned}
\end{equation}
During the evolution the initial discontinuity at $x=0$ 
splits into a shock wave followed by a contact discontinuity, 
both travelling to the right, and a rarefaction wave 
travelling to the left.

\autoref{fig:Sod} depicts our results at time $t = 0.6$ 
for the best behaving high-order WENO5 and low-order 
GODUNOV reconstruction schemes at resolution $\Delta x = 1.25 
\times 10^{-3}$. It is evident from \autoref{fig:Sod} 
that the high-order scheme reproduces all the features 
of the Sod shock-tube quite accurately. A closer examination 
of the plots reveals the existence of small wiggles on 
the horizontal parts between the tail of the rarefaction 
and the shock; see, on the top panel of \autoref{fig:Sod}, 
the inset zoomed-in view of the horizontal portion of 
the velocity profile in question. The maximum amplitude 
of these wiggles is of the order of $\sim 10^{-3}$. 
The use of the low-order scheme prevents the appearance 
of these small wiggles but smears out considerably the 
profiles of the primitive variables, especially around 
the contact discontinuity. However, whichever scheme 
is used (low- or high-order) the oscillations at 
the discontinuities observed in \cite{Guercilena:2016fdl} 
are not present here.

\subsection{Blast waves}

\subsubsection{Blast wave 1}

\begin{figure}[t]
 \centering 
  \includegraphics[width=0.45\textwidth]{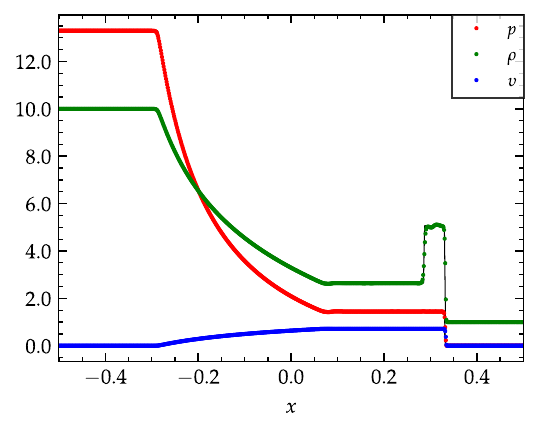}
  \put(-165,180){Blast wave 1: \EFL{-WENO5}}\\
  \vspace{-2mm}
  \includegraphics[width=0.45\textwidth]{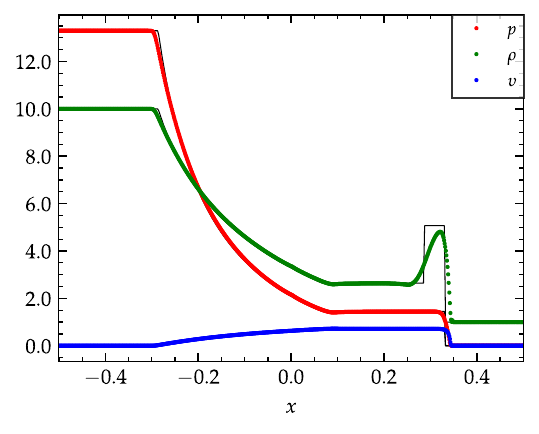}
  \put(-170,180){Blast wave 1: \EFL{-GODUNOV}}
   \caption{Profiles of the rest-mass density (green \green{\textbullet}), 
   velocity (blue \blue{\textbullet}) and pressure (red \red{\textbullet}) 
   for the special-relativistic blast wave 1 test \cite{Marti:1999wi} 
   at $t=0.4$. Top: WENO5 reconstruction. Bottom: GODUNOV reconstruction. 
   The solution is computed on a grid of 800 points with resolution 
   $\Delta x = 1.25 \times 10^{-3}$. Solid black lines are the exact 
   solutions.}
   \label{fig:blast1}
\end{figure}

We continue now with more challenging shock-tube tests. 
We start with the relativistic blast wave 1 test 
described in \cite{Marti:1999wi}. For an ideal EoS 
\eqref{eq:gamma_law} with adiabatic index $\Gamma = 5/3$ 
the initial values of the primitive variables read 
\begin{equation}
 \label{eq:blast1_id}
  \begin{aligned}
   ( p_L, \rho_L, \upsilon_L, \epsilon_L ) &= (13.33, 10, 0, 1.995 ), \\ 
   ( p_R, \rho_R, \upsilon_R, \epsilon_R ) &= (0    , 1 , 0, 0 ).
  \end{aligned}
\end{equation}
The above data is evolved with the RK3 time integrator 
and a CFL factor 0.25 on a numerical grid that spans the 
domain $[-0.5, 0.5]$ along the x-axis. The numerical domain 
is covered with 800 grid-points (resolution $1.25 \times 
10^{-3}$). The numerical solutions at $t = 0.4$ are depicted 
in \autoref{fig:blast1}. The best performing high- and 
low-order schemes for the present shock-tube test are 
the WENO5 and GODUNOV reconstruction schemes, respectively.
Both capture on a quite satisfactory level the main features 
of the exact solutions.
 
\subsubsection{Blast wave 2}

\begin{figure}[t]
 \centering 
  \includegraphics[width=0.45\textwidth]{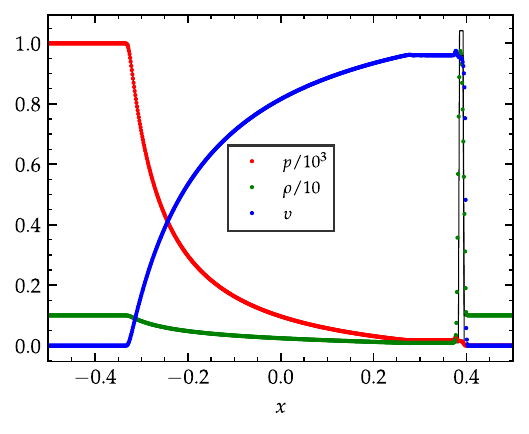}
  \put(-160,185){Blast wave 2: \EFL{-WENO5}}\\
  \vspace{-2mm}
  \includegraphics[width=0.45\textwidth]{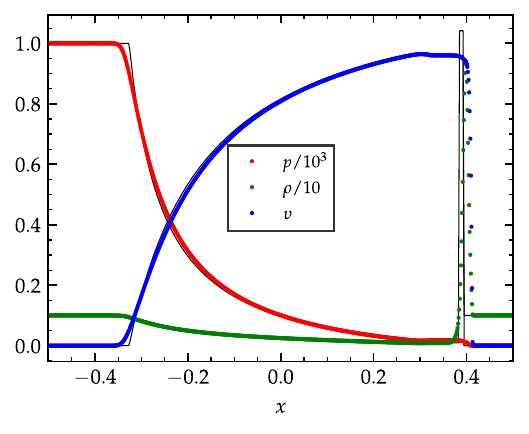}
  \put(-170,185){Blast wave 2: \EFL{-GODUNOV}}
   \caption{Profiles of the rest-mass density (green \green{\textbullet}), 
   velocity (blue \blue{\textbullet}) and pressure (red \red{\textbullet}) 
   for the special-relativistic blast wave 2 test \cite{Marti:1999wi} 
   at $t = 0.4$. Top: WENO5 reconstruction. Bottom: GODUNOV reconstruction. 
   The solution is computed on a grid of 800 points with resolution 
   $\Delta x = 1.25 \times 10^{-3}$. Solid black lines are the exact 
   solutions.}
 \label{fig:blast2}
\end{figure}

\begin{figure*}[t]
  \centering
  \begin{tabular}[c]{ccc}
   \includegraphics[width=0.32\textwidth]{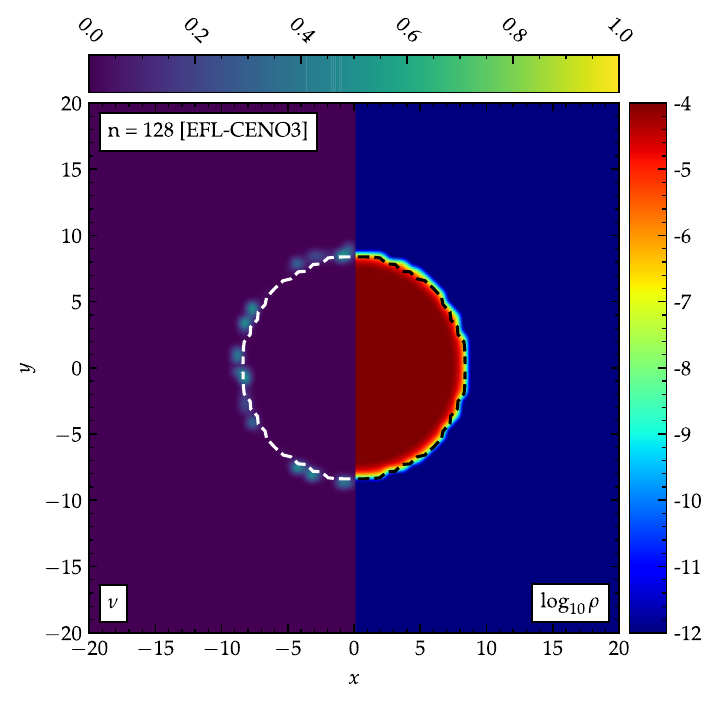}
   \includegraphics[width=0.32\textwidth]{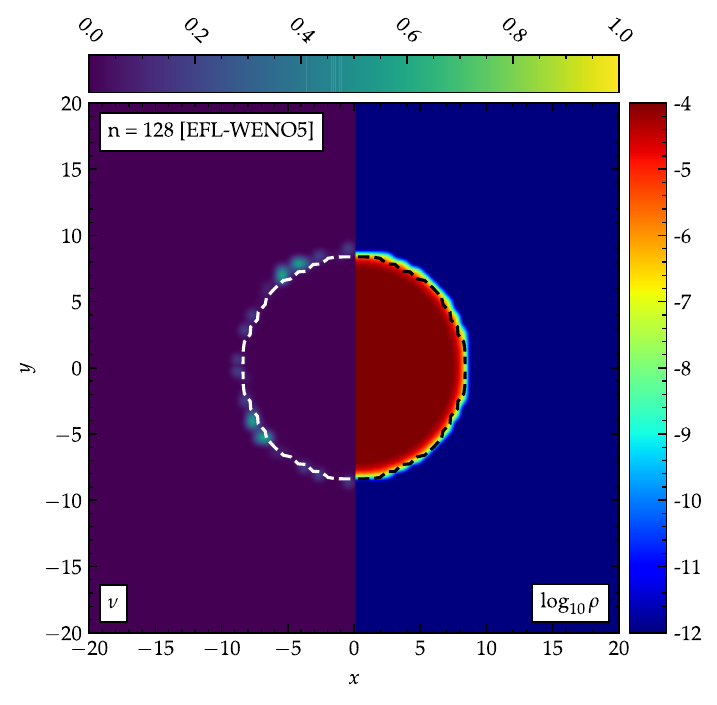}
   \includegraphics[width=0.32\textwidth]{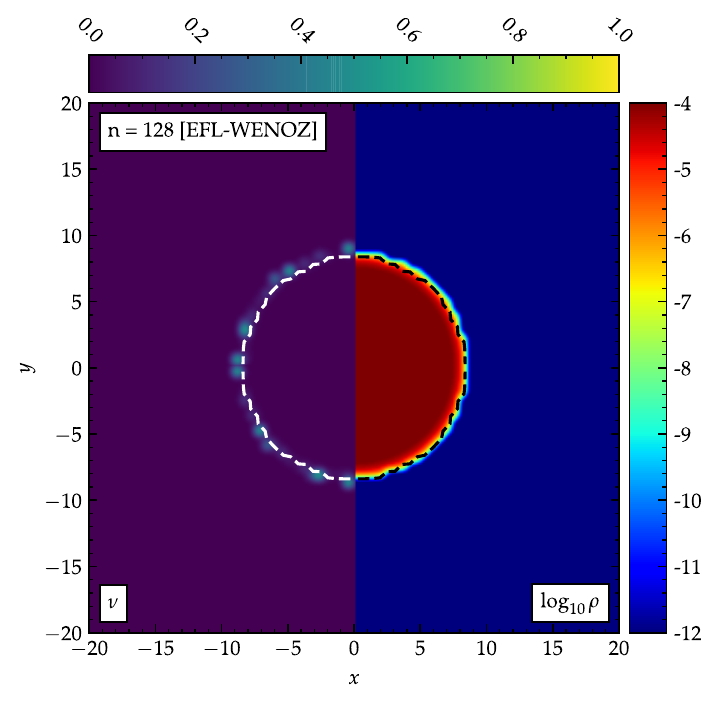}
   \\
   \includegraphics[width=0.32\textwidth]{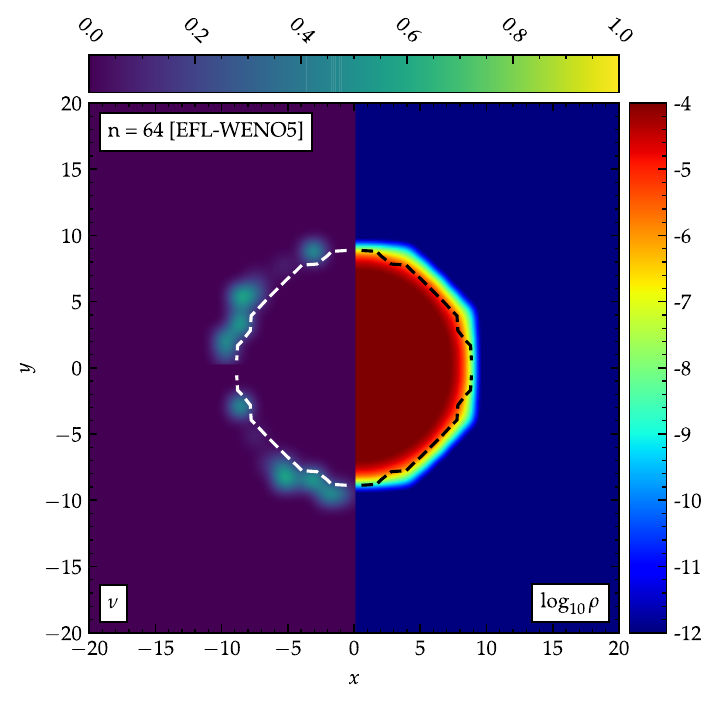}
   \includegraphics[width=0.32\textwidth]{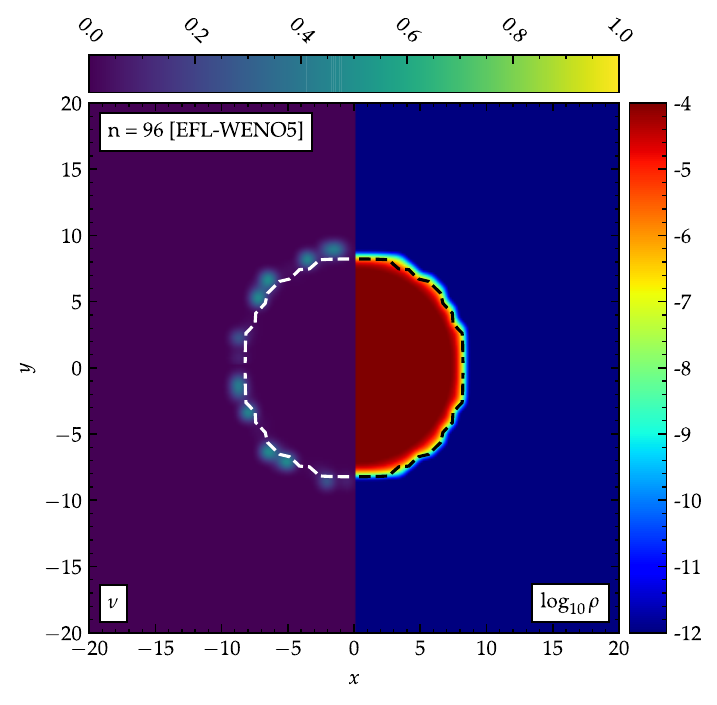}
   \includegraphics[width=0.32\textwidth]{fig05b.pdf}
  \end{tabular}    
  \caption{Two-dimensional profiles of the rest-mass density 
  $\rho$ (right half of the plots) and entropy production function 
  $\nu$ (left half of the plots) for a static TOV star in a dynamical 
  spacetime with $\Gamma$-law EoS at $t=1000$. Top (left to 
  right): CENO3, WENO5 and WENOZ reconstruction schemes with 
  $n=128$. Bottom (left to right): The WENO5 scheme with 
  increasing resolution of 64, 96 and 128 grid points, 
  respectively. The other two schemes show similar behaviour. 
  Dashed lines denote the surface of the NS at $t=0$.}
  \label{fig:ideal_dyn_hybrid_plots}
\end{figure*}

Our final shock-tube test is the blast wave 2 
test \cite{Marti:1999wi}, where the discontinuity 
in the initial data of the pressure and the specific 
energy is of the order of $10^5$. The initial values 
of the primitive variables in this rather extreme 
scenario are the following
\begin{equation}
 \label{eq:blast2_id}
  \begin{aligned}
   ( p_L, \rho_L, \upsilon_L, \epsilon_L ) &= (1000, 1, 0, 1500 ), \\ 
   ( p_R, \rho_R, \upsilon_R, \epsilon_R ) &= (0.01, 1 ,0, 0.015 ).
  \end{aligned}
\end{equation}
We assume again an ideal EoS \eqref{eq:gamma_law} with 
adiabatic index $\Gamma = 5/3$. The numerical solutions 
resulting from the evolution of \eqref{eq:blast2_id} 
are computed on the domain $[-0.5, 0.5]$ (resolution 
$1.25 \times 10^{-3}$) with the RK3 integrator and a 
CFL factor of 0.25. The numerical solutions at $t = 0.4$ 
are presented in \autoref{fig:blast2}. Therein the best 
behaving high- and low-order reconstruction schemes 
are presented and compared to the exact solution. Notice 
the small wiggle appearing on the profiles of the velocity 
and pressure close to the shock. It is definitely not 
an oscillation but some kind of by-product of the EFL 
method as its location coincides with a peak of the 
entropy production function $\nu$. Apart from this feature 
the EFL simulations reproduce to a fairly satisfactory 
degree all the features of the exact solutions.

\section{Single star evolutions}
\label{sec:SNS}
  
Next, in order to test the performance of the EFL method 
in a three-dimensional general relativistic setting, we 
study the evolution of single NS spacetimes. These are 
very challenging tests as the stationarity of the stars 
favours the accumulation and growth of errors, especially 
around the location of the surface where the gradient of 
the hydrodynamical variables experience an abrupt change. 
Unavoidably, the overall accuracy of the simulations is 
affected. At the same time, these tests provide us with 
the exact solution that allows us to study the convergence 
properties of the numerical solutions in detail. We compare 
the performance of the EFL method with different reconstruction 
schemes. Finally, our results are compared with those of 
\cite{Guercilena:2016fdl,Bernuzzi:2016pie}. For comparison 
we use the results obtain with i) a second-order scheme 
(LLF-WENOZ) that uses the LLF scheme for the fluxes and 
WENOZ for primitive reconstruction \cite{Thierfelder:2011yi} 
and ii) a ``hybrid'' algorithm (HO-LLF-WENOZ) that employs 
the high-order HO-WENOZ scheme above a certain density 
threshold $\rho_\mathrm{hyb}$ and switches to the standard 
second-order LLF-WENOZ method below $\rho_\mathrm{hyb}$ 
\cite{Bernuzzi:2016pie}. 

\begin{table}[t]
 \centering   
 \caption{Grid configurations of single star simulations. 
 Columns (left to right): name of simulation, $L$: number 
 of fixed refinement levels, $n$: number of points per 
 direction, $h_{L-1}$: resolution per direction in the 
 finest level $l=L-1$, $h_0$: resolution per direction in 
 the coarsest level $l = 0$.}
   \begin{tabular}{c|cccccc}        
    \hline
    \hline
    Name & $L$ & $n$ & $h_{L-1}$ & $h_0$ \\
    \hline
    \multirow{3}{3em}{TOV} 
    & 3  &  64  & 0.281  & 1.125  \\
    & 3  &  96  & 0.188  & 0.750   \\
    & 3  &  128 & 0.141  & 0.563   \\
    \hline
    \multirow{3}{3em}{RNS} 
    & 3  &  64  & 0.422  & 1.688  \\
    & 3  &  96  & 0.281  & 1.125   \\
    & 3  &  128 & 0.211  & 0.845   \\
    \hline
    \hline
   \end{tabular}  
 \label{tab:SNS_grid}
\end{table} 

In the following, we evolve stable rotating or non-rotating 
neutron stars \cite{Oppenheimer:1939ne} in a dynamically 
evolved spacetime. The NS matter is here described by a 
$\Gamma$-law EoS with $\Gamma=2$. The grid is composed of 
three fixed refinement levels. Simulations are performed 
at resolutions $n=(64,96,128)$ points leading to a grid 
spacing $h$ that depends on the specific setting of the 
NS under investigation. For each NS configuration the 
resolution is explicitly given in \autoref{tab:SNS_grid}. 
It is ensured that the NS is entirely covered by the finest 
box at any given resolution. Radiative (absorbing) boundary 
conditions are used for all single star simulations.

\subsection{TOV star}
\label{sec:static_idea_dyn_tov}

Tolmann-Oppenheimer-Volkoff (TOV) initial data are 
constructed using a $\Gamma=2$ polytrope model with 
gravitational mass $M=1.4$, baryonic mass $M_b=1.506$ 
and central rest-mass density $\rho_c=1.28 \times10^{-3}$. 
The spacetime is dynamically evolved and the BSSNOK 
scheme is used for the evolution of the metric. 

The two-dimensional profiles of the entropy production function 
$\nu$ (left half plane) and rest-mass density (right half plane) 
are depicted on the hybrid plots of \autoref{fig:ideal_dyn_hybrid_plots}.
The three different reconstruction schemes that were used 
here are depicted on the top panel of \autoref{fig:ideal_dyn_hybrid_plots}. 
As expected, a local annular peak of the entropy production 
function $\nu$ is observed around the location of the surface 
of the TOV star. There the gradient of the hydrodynamical 
variables experiences a violent variation which leads to 
the production of large values of $\nu$. The entropy produced 
during the evolution automatically captures the location 
of the star surface. In the interior of the NS the entropy 
production function $\nu$ is as 
expected approximately zero and tends to zero with increasing 
resolution. It is evident from \autoref{fig:ideal_dyn_hybrid_plots} 
that all three reconstruction schemes locate quite accurately 
the surface of the star. In turn, the accurate flagging of 
the surface triggers the use of the stable numerical flux 
around the surface of the star where the hydrodynamical 
variables experience a steep decline. The use of the stable 
scheme in the problematic regions guarantees the stability 
of the star during the evolution. These features of the entropy 
production profile are quite general in all the TOV simulations 
we performed.

\begin{figure}[t]
 \centering 
  \includegraphics[width=0.45\textwidth]{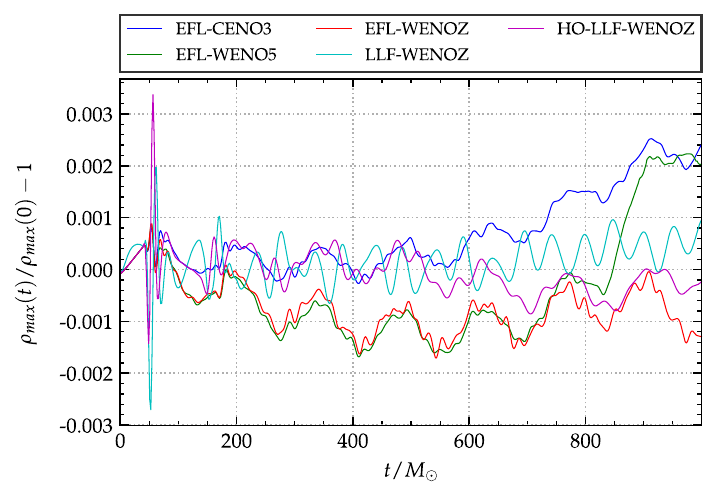}
  \includegraphics[width=0.45\textwidth]{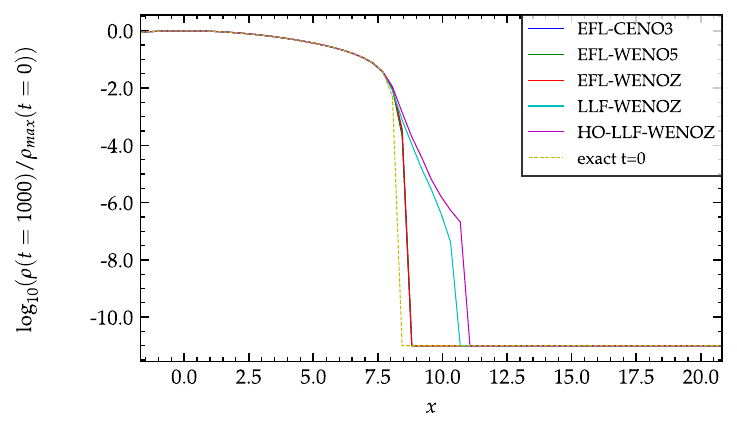}
   \caption{TOV star in a dynamical spacetime with 
   $\Gamma$-law EoS. Top: Central rest-mass density evolution 
   for simulations of a single spherical star with $n=96$ 
   for different EFL reconstruction schemes. Bottom: One-dimensional 
   rest-mass density profiles in the $x$-direction at time 
   $t=1000$ with $n=96$. The profiles in the y- and z-direction 
   are identical.}
   \label{fig:ideal_rho_max_dyn}
\end{figure}

Another very interesting feature of the EFL method, that 
was also stressed in \cite{Guercilena:2016fdl}, is the behaviour 
of the entropy production profile with increasing resolution. 
The bottom panel of \autoref{fig:ideal_dyn_hybrid_plots} 
shows this behaviour: with increasing resolution the entropy 
production function's peaks sharpen and are better localised 
around the surface of the star. This shows that the EFL 
method is able to adjust the entropy production function 
$\nu$ according to the size of the grid-cells. 
The finer they become, the more accurate the problematic 
regions are flagged by the entropy. Ideally, the entropy 
production profile will tend to a delta function located 
around the surface of the star at infinite resolution. 

Having secured the proper flagging of the problematic 
regions and the correct implementation of the EFL method, 
we check further its performance during the evolution of 
the TOV star by monitoring the dynamical behaviour of the 
central rest-mass density. The oscillation of the central 
rest-mass density $\rho_{\rm max}$ of the EFL method with 
different reconstruction schemes is presented, together 
with the LLF-WENOZ and HO-LLF-WENOZ methods \cite{Bernuzzi:2016pie}, 
on the top panel of \autoref{fig:ideal_rho_max_dyn}. The 
performance (i.e.\ the amplitude of the oscillations) of 
the EFL method is comparable to the LLF-WENOZ and HO-LLF-WENOZ 
schemes and to the corresponding results of Fig.~12 in 
\cite{Guercilena:2016fdl}. 

In the bottom panel of \autoref{fig:ideal_rho_max_dyn} the 
profile of the rest-mass density relative to its initial 
maximum value $\rho_{\rm max}$ along the $x$-direction is depicted. 
(The profiles along the $y$- and $z$-direction are, as expected 
from the spherical symmetry of the TOV star, identical.) It 
is evident that the EFL method manages to capture the sharp 
transition between the interior of the TOV star and the outside 
vacuum better than the LLF-WENOZ and HO-LLF-WENOZ schemes. 
It has been also observed that the EFL profiles converge 
to the exact profile with increasing resolution. Comparing 
now our results with the corresponding ones of Fig.~11 in 
\cite{Guercilena:2016fdl}, one can readily conclude that 
the EFL method does not experience the direction dependent 
oscillations reported in \cite{Guercilena:2016fdl} in any 
direction, including the diagonal.

\begin{figure}[t]
 \centering
  \includegraphics[width=0.45\textwidth]{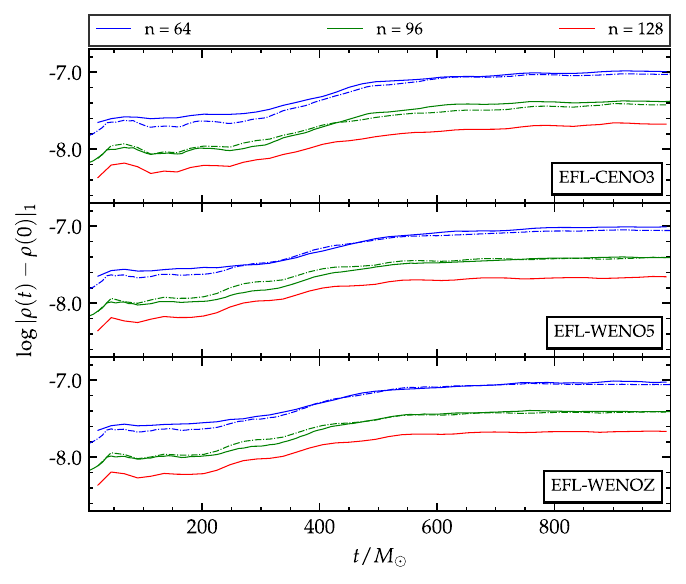} 
   \caption{Evolution of the $L_{1}$-distance $||\rho(t)-\rho(0)||_{1}$ 
   for a static TOV star in a dynamically evolved spacetime 
   with a $\Gamma$-law EoS. All three EFL reconstruction 
   schemes are presented. Dashed lines show results scaled 
   to second-order.} 
 \label{fig:ideal_l1_rho_dyn}
\end{figure}

The static character of the TOV star enables us to check the 
convergence properties of the EFL method as the exact solution 
can be read off from the initial data. Here we consider the 
$L_1$-norm of the difference between the three-dimensional 
evolution profile of the rest-mass density and the corresponding 
exact solution (initial data) and study its behaviour with 
time. The $L_1$-distance from the exact solution for the three 
reconstruction schemes used here are depicted in \autoref{fig:ideal_l1_rho_dyn}. 
The convergence rate of all the schemes considered is approximately 
second order in agreement with the result of \cite{Bernuzzi:2016pie}
and the fact that the error at the stellar surface dominates 
the evolution. Notice though that, for the same resolution, 
the absolute errors of the EFL method are in average 100 times 
smaller than the ones observed in \cite{Bernuzzi:2016pie}. 

\subsection{Rotating neutron star}
\label{sec:rot_ideal_dyn}

\begin{figure}[t]
 \centering 
  \includegraphics[width=0.45\textwidth]{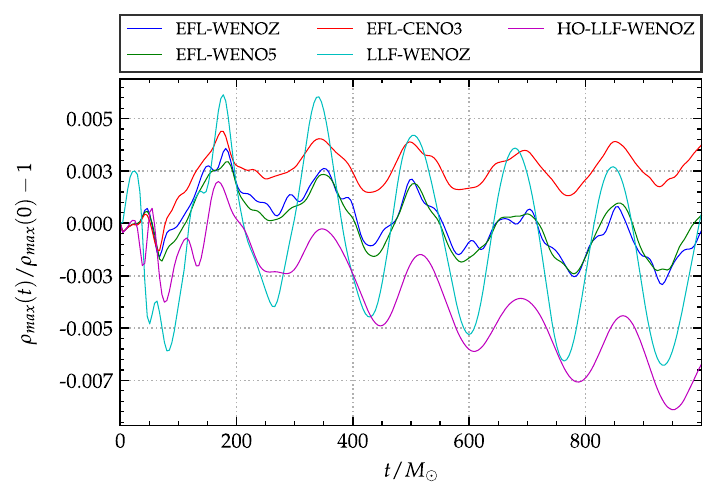}
   \caption{Central rest-mass density evolution for simulations 
   of a stationary RNS in a dynamical spacetime with $\Gamma$-law 
   EoS and $n=96$. Different EFL reconstruction schemes are shown.}
   \label{fig:rho_max_dyn_ideal_rot}
\end{figure}

\begin{figure}[t]
 \centering 
  \includegraphics[width=0.45\textwidth]{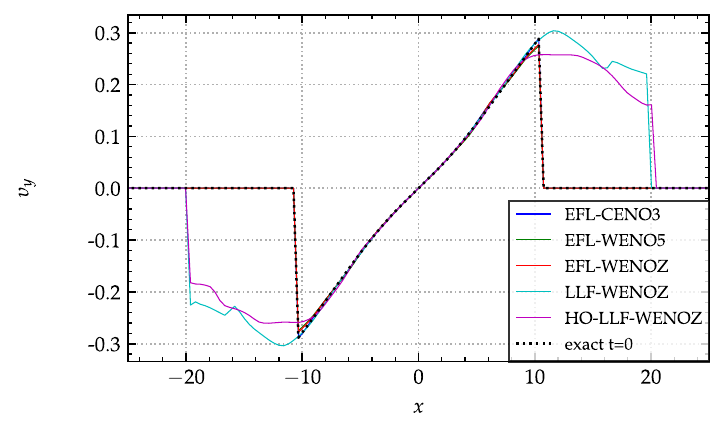}
  \includegraphics[width=0.45\textwidth]{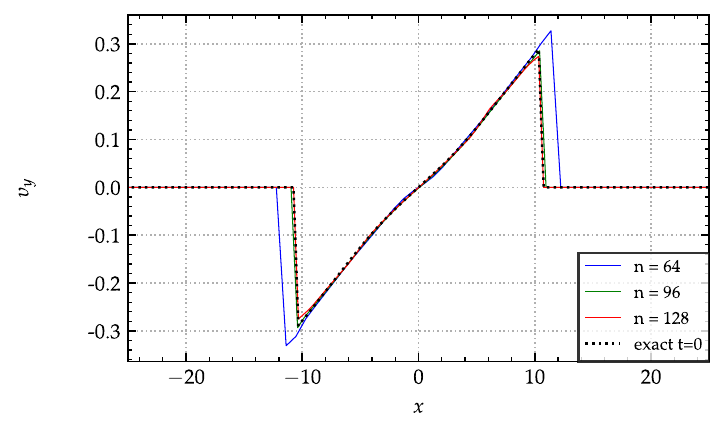}
   \caption{Velocity profile of a stationary rotating neutron 
   star in a dynamical spacetime with $\Gamma$-law EoS. 
   Top: One-dimensional profile of the velocity component 
   $\upsilon_y$ along the $x$-direction at time $t=1000$ 
   (four periods) with $n=128$. Bottom: The $\upsilon_y$ 
   profile of the WENO5 scheme with increasing resolution.}
   \label{fig:vy_profile_dyn_ideal_rot}
\end{figure}

We proceed now to the study of stationary neutron stars. The Rotating 
Neutron Star (RNS) code \cite{Stergioulas:1994ea,Nozawa:1998ak} is 
utilised to construct initial data for a stable uniformly rotating 
neutron star of central rest-mass density $\rho_c=1.28 \times 10^{-3}$, 
axes ratio 0.65 and gravitational mass $M=1.666 M_\odot$ described 
by a polytropic EoS with $\Gamma=2$. This is the BU7 
model described in \cite{Dimmelmeier:2005zk}.

The star is evolved with the $\Gamma$-law EoS and the metric components 
with the Z4c scheme. The spacetime is dynamically evolved. 

We start by checking the behaviour of the central rest-mass 
density with time. \autoref{fig:rho_max_dyn_ideal_rot} 
presents the evolution of the central rest-mass density 
for the EFL method (with three different reconstruction 
schemes) and compares to the LLF-WENOZ and HO-LLF-WENOZ 
methods \cite{Bernuzzi:2016pie}. The resulting oscillating 
behaviour is triggered by atmosphere effects and converges 
to 0 with increasing resolution. The results of all three 
methods are comparable with the oscillatory behaviour of 
the EFL method to be the smallest. 

\begin{figure}[t]
 \centering
  \includegraphics[width=0.45\textwidth]{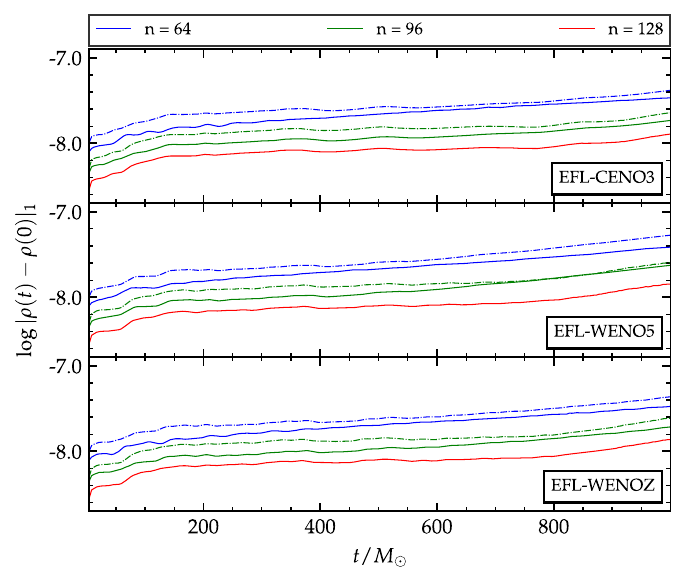} 
   \caption{Evolution of the $L_{1}$-distance $||\rho(t)-\rho(0)||_{1}$ 
   for a rotating neutron star in a dynamically evolved spacetime 
   with $\Gamma$-law EoS.  Dashed lines show results scaled to 
   second-order.} 
 \label{fig:ideal_l1_rho_dyn_rot}
\end{figure}

For uniformly rotating neutron stars one expects that the 
velocity increases linearly with the distance from the 
centre of the star, reaches its maximum value at the 
surface and drops to zero from there on. The top panel of 
\autoref{fig:vy_profile_dyn_ideal_rot} shows the velocity 
component $\upsilon_y$ along the $x$-axis after four 
rotational periods (4P). The results for the EFL method 
(with three different reconstruction schemes) are presented 
and compared to the LLF-WENOZ and HO-LLF-WENOZ methods. 
It is apparent that the EFL method is superior in preserving 
the original velocity profile of the star (with the CENO3 
scheme capturing the initial profile exactly). Comparing 
our results with the ones that can be found in the literature 
\cite{Font:2001ew,Stergioulas:2000,Font:1999wh}, we observe 
that the EFL method can capture better the rapid transition 
from the maximum value of the velocity at the surface 
of the star to zero just outside it. At the bottom panel 
of \autoref{fig:vy_profile_dyn_ideal_rot} the profile 
of $\upsilon_y$ for the WENO5 scheme is presented with 
increasing resolution---the other two reconstruction 
schemes show similar behaviour. The numerical solutions 
converge to the initial exact velocity profile with 
increasing resolution. The above results clearly demonstrate 
the ability of the EFL method to maintain the initial 
stationary equilibrium configuration during the evolution. 

Finally, we check the convergence properties of the 
$L_{1}$-distance of $\rho(t)$ from the exact solution $\rho(0)$. 
\autoref{fig:ideal_l1_rho_dyn_rot} depicts the time-evolution 
of the $L_1$-norm of the difference between the three-dimensional 
profile of the rest-mass density and its initial profile for 
the three reconstruction schemes used here. All schemes show 
approximately second-order convergence.

\section{Binary neutron star evolutions}
\label{sec:BNS}

\subsection{Initial data and numerical setup}

Having thoroughly tested the EFL method with several 
special relativistic and single NS configurations, we 
move to discuss the EFL method in general relativistic 
simulations of neutron star binaries. In the following, 
we study the dynamics of two specific BNS configurations: 
\BAM{100} and \BAM{97}, see \cite{Dietrich:2018phi}. 
We chose these two BNS simulations because they enable 
us to study the performance of the EFL method in a short 
(\BAM{100}) and a long (\BAM{97}) BNS dynamical evolution. 
The neutron stars merge after approximately 3 revolutions 
for \BAM{100} and after 10 for \BAM{97}. In addition, 
both simulations were already extensively studied 
in the literature, see \cite{Bernuzzi:2016pie}, to which 
we refer and compare our results with. In \autoref{sec:BNS_hybrid_plots} 
yet another three-orbit simulation, \BAM{27}, is used in 
order the test the \EFL{} method with a different EoS. 
Although, our results for \BAM{27} are consistent with 
the ones presented here, in the following, for the sake 
of presentational clarity, we do not discuss \BAM{27} but 
focus on the other two simulations of \autoref{tab:BNS_sim}.

\begin{table}[t]
 \centering    
 \caption{BNS quasicircular initial data. Columns: name, EoS, 
 number of orbits, binary mass $M$, rest-mass $M_b$, ADM mass 
 $M_\mathrm{ADM}$, angular momentum $J_0$, GW frequency $2M\Omega_0$. 
 All configurations are equal-masses and irrotational.}
   \begin{tabular}{ccccccccc}        
    \hline
    \hline
    Name & ID & EoS & orbits & $M$ & $M_b$ & $M_\mathrm{ADM}$ & $J_0$ & $2M\Omega_0$\\
    \hline
    \BAM{100} & Lorene & SLy & 3  & 2.700 & 2.989 & 2.671 & 6.872 & 0.060 \\
    \BAM{27}  & Lorene & $\Gamma=2$  & 3  & 3.030 & 3.250 & 2.998 & 8.835 & 0.055  \\
    \BAM{97}  & Lorene & SLy & 10 & 2.700 & 2.989 & 2.678 & 7.658 & 0.038 \\
    \hline
    \hline
   \end{tabular}
 \label{tab:BNS_sim}
\end{table} 

\begin{table}[t]
 \centering    
 \caption{Grid configurations of BNS simulations. Columns 
 (left to right): name of BNS simulation, $L$: refinement 
 levels, $l^\mathrm{mv}$: minimum moving level index, 
 $n^\mathrm{fix}$: number of points per direction in 
 fixed levels, $n$: number of points per direction in 
 moving  levels, $h_{L-1}$: resolution per direction 
 in the finest level $l=L-1$, $h_0$: resolution per 
 direction in the coarsest level $l = 0$.}
   \begin{tabular}{c|cccccc}        
    \hline
    \hline
    Name & $L$ & $l^\mathrm{mv}$ & $n^\mathrm{fix}$ & $n$ & $h_{L-1}$ & $h_0$ \\
    \hline
    \multirow{4}{4em}{\BAM{100}} 
    & 7 & 2 & 128 & 64  & 0.228  & 14.592  \\
    & 7 & 2 & 192 & 96  & 0.152  & 9.728   \\
    & 7 & 2 & 256 & 128 & 0.114  & 7.296   \\
    & 7 & 2 & 320 & 160 & 0.0912 & 5.8368  \\
    \hline
    \multirow{4}{4em}{\BAM{97}} 
    & 7 & 2 & 160 & 64  & 0.228  & 14.592  \\
    & 7 & 2 & 240 & 96  & 0.152  & 9.728   \\
    & 7 & 2 & 320 & 128 & 0.114  & 7.296   \\
    & 7 & 2 & 400 & 160 & 0.0912 & 5.8368  \\
    \hline 
    \BAM{27}\footnote{\justify \vspace{-6mm} \quad We use 
    only one resolution for \BAM{27} as in the present work 
    we do not present a convergence analysis for it, but 
    just the two-dimensional entropy production profiles 
    of \autoref{fig:hybrid_plots_gamma}.}
    & 7 & 1 & 96 & 64  & 0.312  & 20.0  \\
    \hline
    \hline
   \end{tabular}
 \label{tab:BNS_grid}
\end{table} 

The initial data that we evolved can be found 
in \autoref{tab:BNS_sim}. They are conformally flat 
irrotational BNS configurations in quasicircular orbits 
computed with the Lorene library \cite{Gourgoulhon:2000nn} 
and characterised by the Arnowitt-Deser-Misner (ADM) 
mass-energy $M_{\rm ADM}$, the angular momentum $J_0$, 
the baryonic mass $M_b$ and the dimensionless GW circular 
frequency $M \omega_0$. 

The initial data for \BAM{100} and \BAM{97} were evolved 
with the \EFL{} method in 16 different resolution and reconstruction 
combinations. \BAM{100} was evolved with the CENO3, 
WENO5 and WENOZ reconstructions. For each reconstruction 
four different grid resolutions were considered. The grid 
specifications for all the runs are reported in \autoref{tab:BNS_grid}.
\BAM{97} was evolved only with the WENOZ reconstruction 
scheme. The reason for this is, as it will become apparent 
in the following, that the \BAM{100} results strongly indicate 
that the best performing reconstruction scheme is WENOZ. 
The atmosphere setting for both simulations are $f_{\rm atm}=10^{-11}$ 
and  $f_{\rm thr}=10^2$. The metric is evolved with the 
Z4c scheme. Standard radiative boundary conditions 
are used for all BNS simulations.

\subsection{Qualitative behaviour of the entropy production}
\label{sec:BNS_hybrid_plots}

\begin{figure}[t]
 \centering 
  \includegraphics[width=0.5\textwidth]{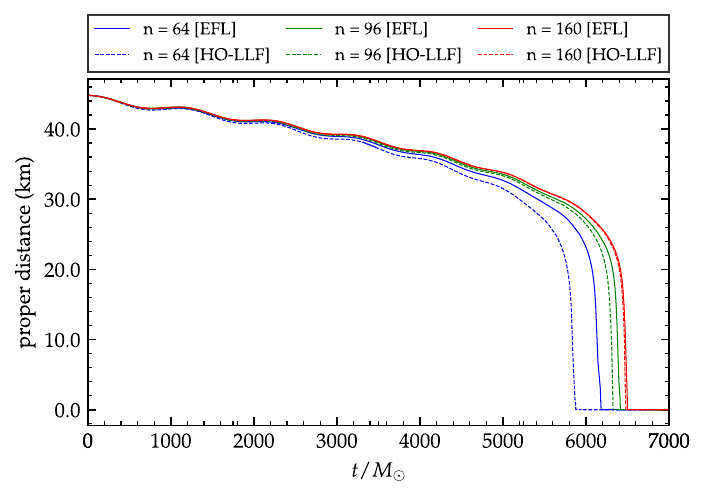}
  \caption{Time evolution of the proper distance between 
  the NSs for the ten-orbit \BAM{97} simulation. Different 
  grid resolutions are presented for the \EFL{-WENOZ} 
  method (solid lines) and  HO-LLF-WENOZ method (dashed 
  lines).}
  \label{fig:proper_d_bam97}
\end{figure}

The most basic and simple check of the \EFL{} method is 
to inspect if the produced NS trajectories agree with 
the ones in the literature \cite{Bernuzzi:2016pie}. 
\autoref{fig:proper_d_bam97} depicts the behaviour 
with time of the proper distance between the NSs of 
the ten-orbit \BAM{97} simulation for the \EFL{-WENOZ} 
and HO-LLF-WENOZ schemes at different resolutions. 
The time of merger can be determined from the vanishing 
of the proper distance. Notice that while for high 
resolutions the two methods agree, for low resolutions 
their behaviour differs as shorter inspirals are expected 
for lower resolutions because of numerical dissipation 
\cite{Bernuzzi:2011aq,Bernuzzi:2012ci}. 
It is evident from \autoref{fig:proper_d_bam97} that 
although the trajectories of the  HO-LLF-WENOZ scheme 
for low resolutions are less accurate than the ones of 
the EFL-WENOZ scheme, they catch up with increasing 
resolution. Thus, one would expect that the trajectories 
of the HO-LLF-WENOZ scheme converge faster to the actual 
trajectory of the inspiraling NSs. Indeed, by conducting 
self-convergence tests for the triplets $n=(64,96,128)$ 
and $n=(96,128,160)$ we conclude that the actual convergence 
rate of the proper distance for the \EFL{-WENOZ} scheme 
is approximately third- and fourth-order, respectively. 
A similar analysis for the HO-LLF-WENOZ scheme shows 
that the convergence rates for the above triplets are 
approximately fourth- and six-order, respectively. 
The trajectories of the three-orbit \BAM{100} simulation 
show similar behaviour. 

The entropy production function $\nu$ plays a central role 
in our method. Hence, it is of great interest to study 
its behaviour during the evolution of BNS merger simulations. 
In the following, we discuss the two-dimensional entropy 
production profiles of two different three-orbit simulations. 
Together with the three-orbit \BAM{100} simulation, we  
present here another three-orbit simulation with a different 
EoS. The reason for this is to exemplify the dependence 
of our method on the EoS used, which is best depicted by 
the entropy production profile. We use the \BAM{27} 
simulation \cite{Dietrich:2018phi} with initial data
parameters given in \autoref{tab:BNS_sim}. 

\begin{figure*}[t]
  \centering 
  \begin{tabular}[c]{ccc}
  \includegraphics[width=0.32\textwidth]{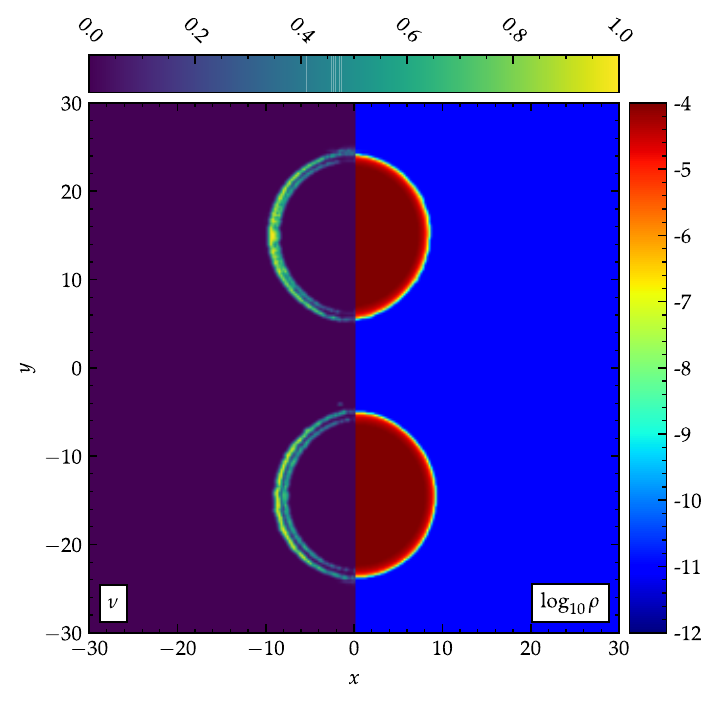}
  \includegraphics[width=0.32\textwidth]{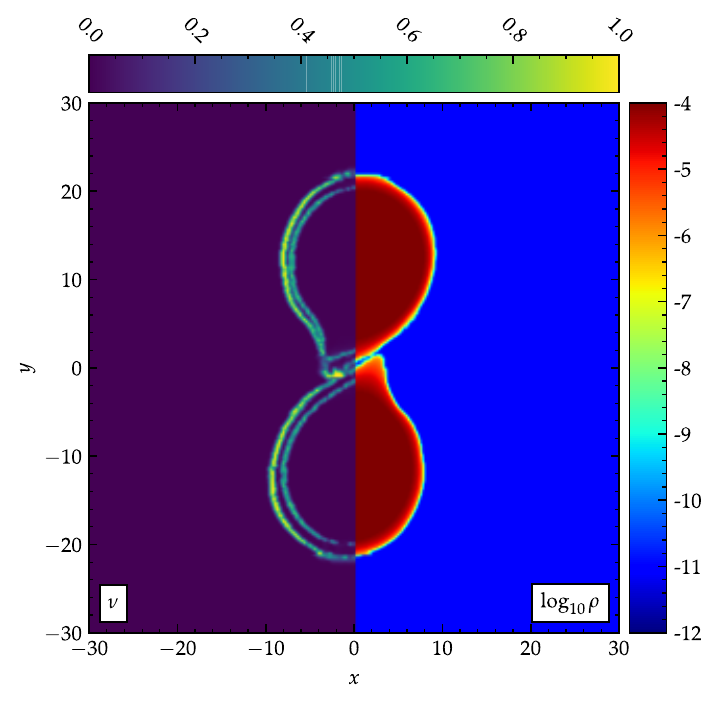}
  \includegraphics[width=0.32\textwidth]{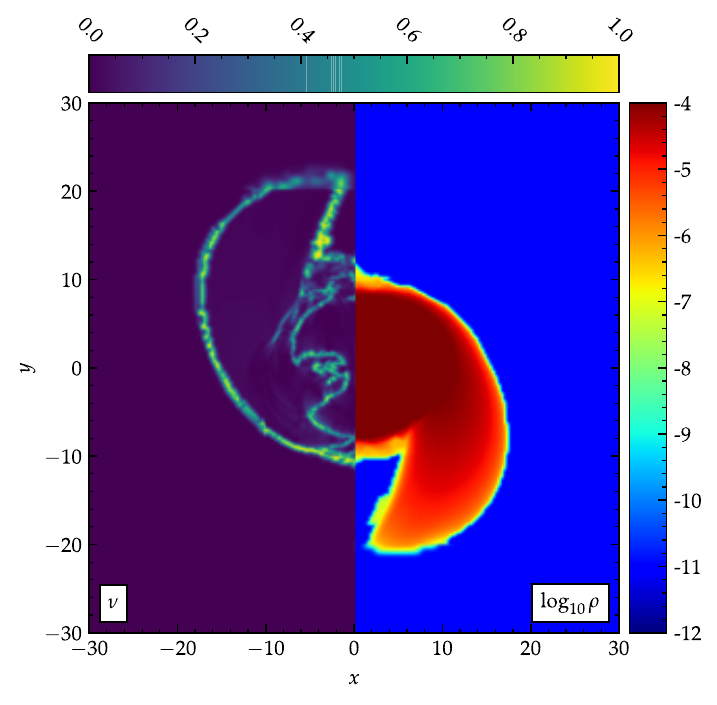}
  \end{tabular}
  \caption{Two-dimensional hybrid plots depicting the entropy 
  production and rest-mass density profiles across different stages 
  of the \BAM{27} simulation. Notice that \BAM{27} uses a $\Gamma$-law 
  EoS. Left to right: inspiral, merger, post-merger.}
  \label{fig:hybrid_plots_gamma}
  \begin{tabular}[c]{ccc}
  \includegraphics[width=0.32\textwidth]{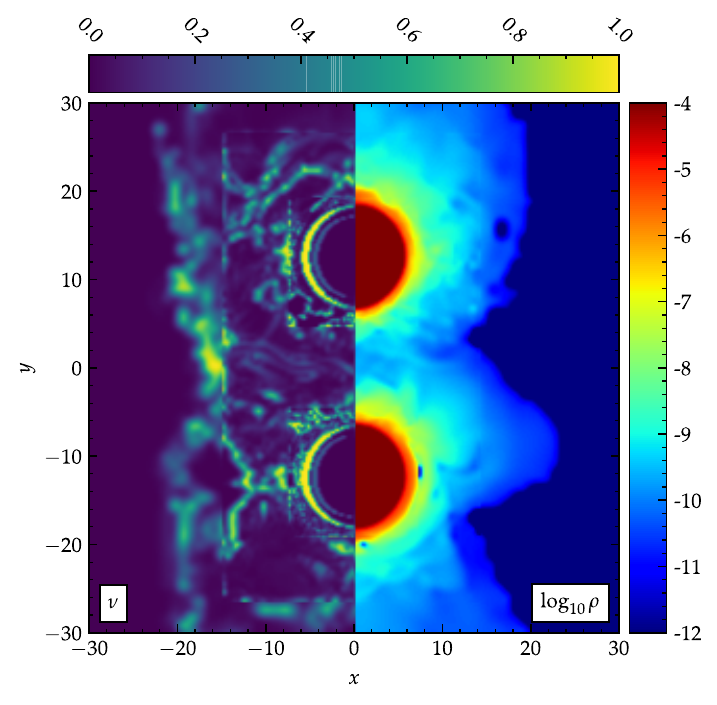}
  \includegraphics[width=0.32\textwidth]{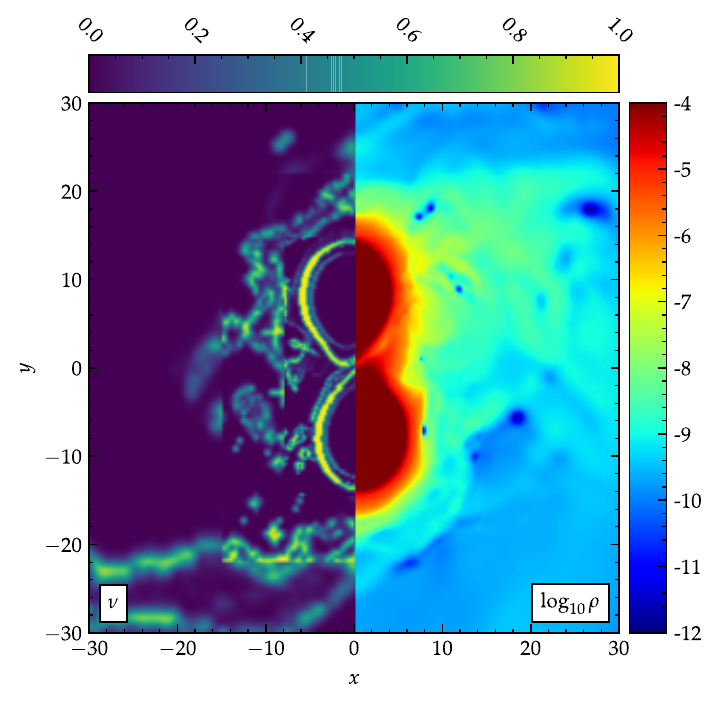}
  \includegraphics[width=0.32\textwidth]{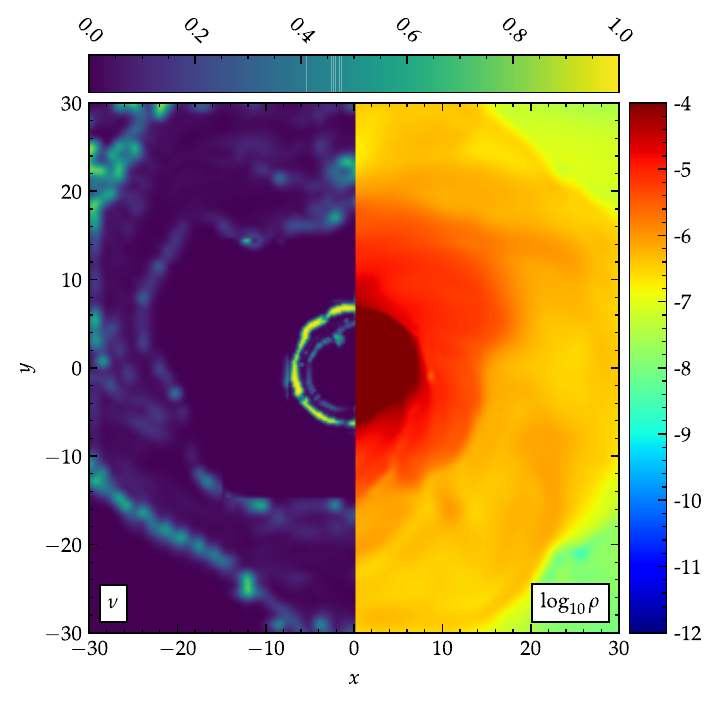}
  \end{tabular}
  \caption{Two-dimensional hybrid plots depicting the entropy 
  production and rest-mass density profiles across different stages 
  of the \BAM{100} simulation. Notice that \BAM{100} uses a SLy EoS. 
  Left to right: inspiral, merger, post-merger.}
  \label{fig:hybrid_plots_sly}
\end{figure*}

In \autoref{fig:hybrid_plots_gamma} and \autoref{fig:hybrid_plots_sly}
we present two-dimensional hybrid plots depicting the entropy 
production function $\nu$ and rest-mass density $\rho$ profiles 
at different stages of a $\Gamma$-law (\BAM{27}) and SLy (\BAM{100}) 
simulation, respectively. The different panels show (from left 
to right) selected snapshots of the inspiral, merger and post-merger 
stages, respectively. We first notice that the SLy simulation 
displays considerably more features in both $\nu$ and $\rho$ 
profiles compared to the $\Gamma$-law simulation. Taking a 
closer look at the $\nu$ profile, we see that during the 
$\Gamma$-law simulation the EFL is only activated at the 
surface of each NS. The same can be also observed 
in the SLy simulation, however, here also regions in exterior 
of the NSs are flagged for limiting. Judging from the corresponding 
rest-mass density plots, the flagging in the exterior is due 
to the SLy simulation carrying a matter cloud around the stars, 
a feature that is absent from the $\Gamma$-law simulation.

During the inspiral phase of \autoref{fig:hybrid_plots_gamma}
we can see that there are actually two concentric layers 
where the EFL is activated around the surface (left panels).
When the stars first touch, their interior and exterior 
layers start merging with each other (middle panels). After 
the first contact the inner layer also starts moving inwards 
before it becomes again concentric with the surface layer 
(right panels). The surface layer continues tracking the 
star's surface during merger which is evident by its 
alignment with the matter-vacuum interface that can be 
seen in the rest-mass density profile. It seems that this 
\textit{double-layer} formation is universal, because 
we find similar behaviour during the RNS evolutions. 
% Most probably the inner layer is triggered by the rotation.
Interestingly, the mass density plots do not show apparent 
features that would need shock treatment in the region where 
the second layer appears. There are two reasons causing this 
double-layering: i) Low resolution: With increasing resolution 
the entropy production function $\nu$ gets better localised 
around the surface of the NSs, see bottom panel of 
\autoref{fig:ideal_dyn_hybrid_plots}, and consequently 
the amplitude of the inner peak of $\nu$ decreases; ii) 
The entropy production function $\nu$ is overproduced by 
setting $c_E=1$ in \eqref{eq:nu_E}. Recall that for the 
sake for generality and simplicity we set $c_E=1$ in all 
our simulations. By choosing a smaller value for the tunable 
constant $c_E$, the values of $\nu$ would scale down accordingly. 
The inner layer then would reduce. While it is possible to 
experiment with the values of $c_E$ to minimize this effect, 
we find that the convergence properties of the solutions 
are not affected by it.

Lastly, the same double layer formation can also be observed 
in \autoref{fig:hybrid_plots_sly}, although the amplitude 
of the inner layer is smaller and the exterior layer appears 
to be wider.

\subsection{Conserved quantities}
\label{sec:BNS_conserved_quant}

\begin{figure}[t]
 \centering 
  \includegraphics[clip,width=0.45\textwidth]{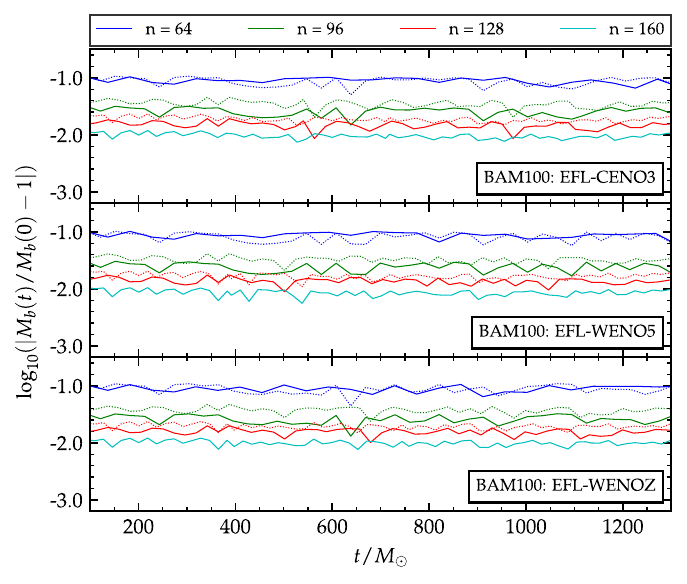}
  \includegraphics[clip,width=0.47\textwidth]{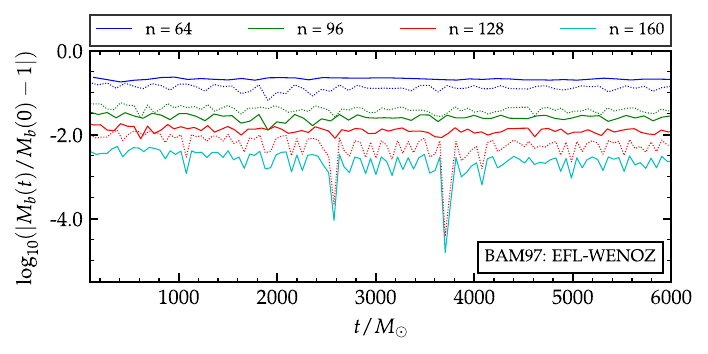}
  \hspace{-3mm}
  \caption{Conservation and convergence of the total 
  rest-mass on the refinement level $l=2$. Top: Three-orbit 
  \BAM{100} simulation. All three reconstruction schemes 
  are presented for different grid resolutions. Dotted 
  lines show results scaled to third-order. Bottom: 
  Ten-orbit \BAM{97} simulation. The WENOZ reconstruction 
  schemes is presented for different grid resolutions. 
  Dotted lines show results scaled to fourth-order.}
  \label{fig:mass_conserv_bns}
\end{figure}

\begin{figure}[t]
 \centering 
  \includegraphics[clip,width=0.45\textwidth]{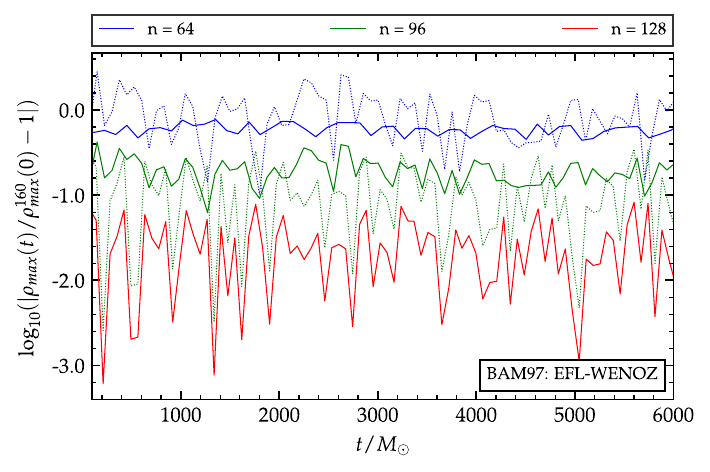}
  \caption{Evolution and convergence of the central rest-mass 
  density on the refinement level $l=1$ for the ten-orbit \BAM{97} 
  simulation. Dotted lines show results scaled to fifth-order. 
  }
  \label{fig:bns_rho_max}
\end{figure}

Conserved quantities can be used as quantitative and 
qualitative diagnostics of the performance of a numerical 
scheme. Therefore, before discussing the waveform accuracy 
of our simulations, we study the convergence properties 
of these quantities. During the BNS evolution we monitor:

\begin{itemize}[leftmargin=*,noitemsep]
\item[i)] The total rest-mass of the matter, 
$M_b=\int d^3 x \sqrt{\gamma}\,D$, where the integral is 
performed over the whole computational domain. The continuity 
equation \eqref{eq:conserv_laws} guarantees that the total 
rest-mass should be conserved in the absence of a net influx 
or outflow of matter. We use a conservative numerical scheme 
\eqref{eq:conserv_PDE}, which is expected to preserve the 
rest-mass to its initial value. This requirement is trivially 
satisfied on a single grid, but violations are generically 
expected in the presence of the artificial atmosphere and 
when adaptive mesh refinement is used, see e.g.~\cite{Dietrich:2015iva}.

\item[ii)] The dynamical behaviour of the central rest-mass 
density, $\rho_{\rm max}(t)$, of the NSs. Unlike the 
stationary single star simulations, this quantity is 
not exactly conserved in BNS simulations because of the 
presence of tidal interactions. However, during the early 
orbits of the inspiral, tidal interactions are weak and 
contribute only small oscillations around the initial 
value. For the considered resolutions, the latter are 
actually smaller than the oscillations induced by truncation 
errors and should converge to zero with increasing resolution. 
Hence, the ratio of the central rest-mass density to its 
initial value for the highest resolution, i.e. the quantity 
$\rho_{\rm max}(t)/\rho^{n_{\rm max}}_{\rm max}(0)$, should 
tend to one with increasing resolution.
  
\item[iii)] The $L_2$ norm of the Hamiltonian constraint. 
It is expected that in the continuum limit it vanishes, 
thus the Hamiltonian constraint of any numerical solution 
must convergence to zero in order to be consistent with 
Einstein's equations.
\end{itemize}  

\begin{figure}[t]
 \centering 
  \includegraphics[clip,width=0.45\textwidth]{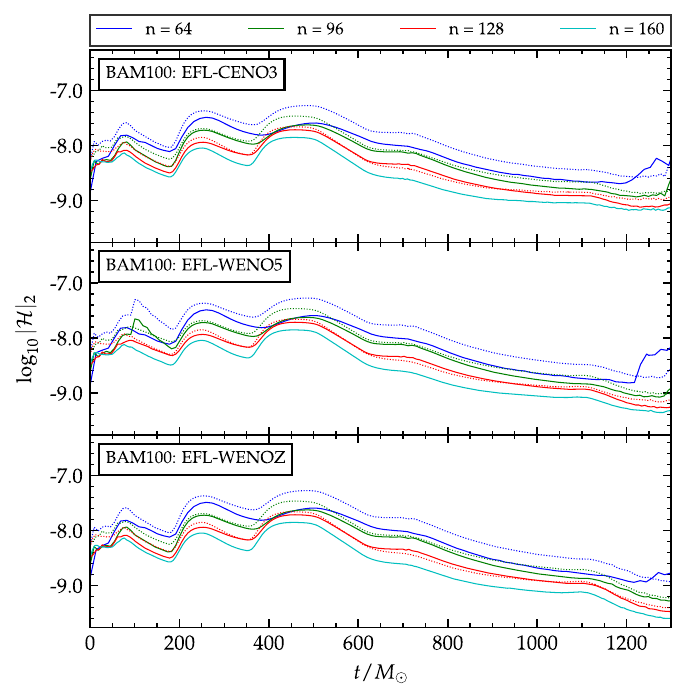}
  \includegraphics[clip,width=0.47\textwidth]{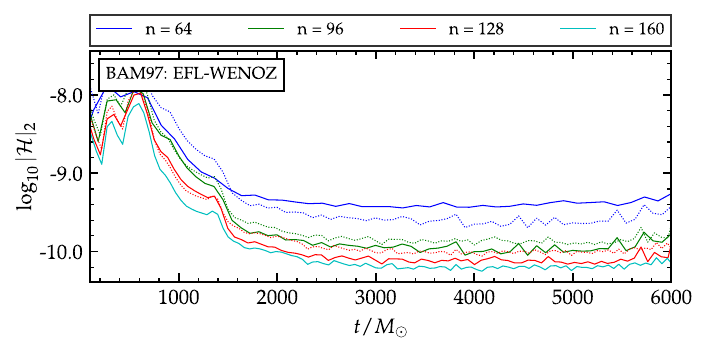}
  \hspace{-3mm}
  \caption{$L_2$ norm of the Hamiltonian constraint on 
  the refinement level $l=1$ for the BNS simulations. Top: 
  Three-orbit \BAM{100} simulation. All three reconstruction 
  schemes are shown. Bottom: Ten-orbit \BAM{97} simulation. 
  Dotted lines show results scaled to second-order.}
  \label{fig:Ham_norm_bns}
\end{figure}
 
\autoref{fig:mass_conserv_bns} depicts the violation of 
the total rest-mass conservation during the inspiral and 
up to merger for the BNS simulations considered here. On 
the top panel the results for the three-orbit \BAM{100} 
simulation are shown. Each sub-panel depicts one of the 
reconstruction schemes used here for four different 
resolutions. The violation of the rest-mass conservation 
during the inspiral phase is mainly caused due to 
artificial atmosphere treatment and mesh refinement 
boundaries. According to \autoref{fig:mass_conserv_bns}, 
during the evolution the mass violation shows small 
oscillations around its initial value, but is neither 
increasing nor decreasing. The mass violation converges 
to zero in an approximately third-order convergence 
pattern with increasing resolution. (Dotted lines show 
results scaled to third-order.) After the merger mass 
loss is caused by the ejected material which decompresses 
while it leaves the central region of the numerical domain 
(not shown in the plot). The performance of all three 
reconstruction schemes is comparable. In the bottom panel 
the respective mass violation for the \BAM{97} simulation 
with the WENOZ scheme is presented. The behaviour of 
the mass violation of \BAM{97} is quantitatively similar 
to the \BAM{100} simulation, although \BAM{97} converges 
with fourth-order and the absolute mass violation is 
smaller at the highest resolution.

The evolution of the central rest-mass density together 
with its convergence pattern on the $l=1$ refinement 
level for the ten-orbit \BAM{97} simulation are presented 
in \autoref{fig:bns_rho_max}. The relative error 
$\rho_{max}(t)/\rho^{160}_{max}(0)$ is used to monitor 
the central rest-mass density during the evolution, 
where $\rho^{160}_{max}(0)$ is the initial value of 
the central rest-mass density for the highest resolution 
$n=160$ used here. It is evident from \autoref{fig:bns_rho_max} 
that the residual $|\rho_{max}(t)/\rho^{160}_{max}(0)-1|$ 
with increasing resolution tends to zero with an approximate 
fifth-order convergence rate. Notice that the observed 
oscillatory behaviour gradually dies out with increasing 
resolution, but because of the logarithmic scale of 
\autoref{fig:bns_rho_max} this feature is not easily 
seen.

\begin{figure*}[t]
  \includegraphics[width=0.95\textwidth]{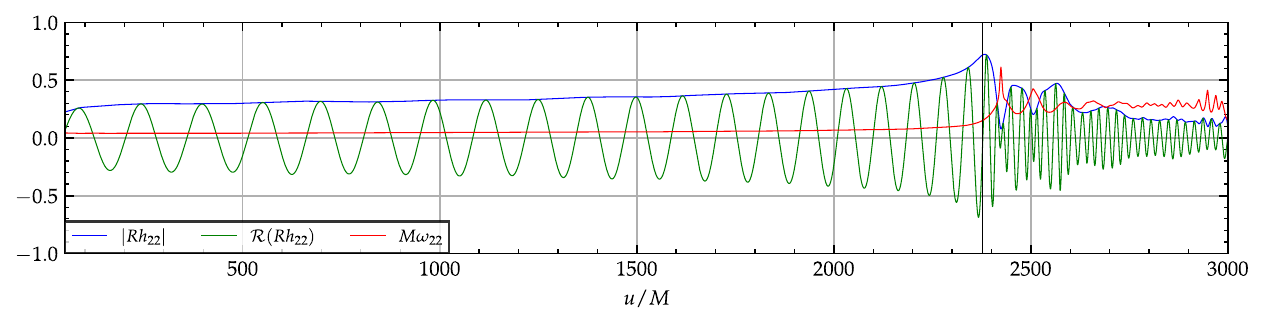}
  \caption{Amplitude (blue) and real part (green) of strain 
  $R h_{22}$ as well as instantaneous frequency $M\omega_{22}$ 
  (red) of the GW signal obtained from the ten-orbit \BAM{97} 
  simulation using the \EFL{-WENOZ} scheme. The time of merger 
  is defined as the first peak in the amplitude $A_{22}$ and 
  is indicated by a black vertical line at $u_{\rm mrg} \approx 2400 M$.}
  \label{fig:gw_bam97}
\end{figure*}

The $L_2$ norm of the Hamiltonian constraint on refinement 
level $l=1$ up to merger with increasing resolution for 
both BNS simulations is presented in \autoref{fig:Ham_norm_bns}. 
In the top panel results for all three reconstruction 
schemes used in the \BAM{100} simulations are presented 
for four different resolutions. In all cases the violation 
of the constraint is of the order of $\sim 10^{-8}$ for 
the lowest resolution of 64 points and decreases to zero 
with increasing resolution. The observed convergence is 
approximately second-order and agrees with the corresponding 
results in \cite{Bernuzzi:2016pie}. (Dotted lines show 
results scaled to second-order convergence.) We attribute 
this behaviour to the to the constraint propagation and 
damping properties of the Z4c evolution system \cite{Weyhausen:2011cg}.
Notice that in all cases during the evolution the constraint 
violation remains below its initial value and only increases, 
as expected, close to merger. Moreover, the Hamiltonian 
violation and convergence is similar in all the AMR levels 
independently on the fact that the matter is well resolved 
or not on the grid. This suggests that truncation error 
from AMR boundaries, boundary conditions and Berger-Oliger 
time interpolation are likely dominant in this quantity 
for Z4c. The performance of all reconstruction schemes is 
comparable, with the WENOZ scheme showing smaller constraint 
violation for the same resolution than the other two schemes. 
In the bottom panel results for the ten-orbit \BAM{97} simulation 
are shown with the WENOZ scheme. The behaviour of the constraint 
violation is similar to the one observed for the \BAM{100} 
simulation, with the slight difference that now the plots 
are a bit less smooth and that the constraint violation is 
smaller for the same resolution. 

This second-order convergence of the violation of the 
Hamiltonian constraint is in stark contrast with the 
observed fourth-order or higher convergence of i) the 
rest conserved quantities studied in the present section, 
ii) the proper distance, see \autoref{sec:BNS_hybrid_plots} 
and iii) the GW phase differences, see \autoref{fig:gw_phase_errors_bam97}. 
Second-order convergence in the violation of the Hamiltonian 
constraint has been observed in all the BNS simulations 
produced with BAM to date, see \cite{Bernuzzi:2011aq,
Thierfelder:2011yi,Hilditch:2012fp,Bernuzzi:2016pie}. 
In addition, the results of \autoref{fig:Ham_norm_bns} 
are similar to the corresponding ones in \cite{Bernuzzi:2016pie}. 
The lower convergence rate in the Hamiltonian constraint
violation is due to the details of BAM's infrastructure, 
and not to the EFL itself. The main reasons are i) the 
fact that for efficiency the primitives are not synchronised 
in BAM and consequently the Hamiltonian is computed from 
the rest-mass density $\rho$ from a half time-step before 
ii) the propagation properties of Z4c, which means that 
an error contribution also comes from the boundary/AMR 
interpolation.

\subsection{Gravitational wave analysis}
\label{sec:gw_analysis}

We discuss here the impact of the \EFL{} scheme on the gravitational 
waveforms. We follow closely \cite{Bernuzzi:2016pie} and examine
the phase convergence in the inspiral-merger GWs and the associated 
error budget.

GWs are computed from the curvature scalar field $\Psi_4$ on 
coordinate spheres that are a distance $r$ from the origin of 
the computational domain. GW reconstruction is done by expanding 
$\Psi_4$ into spin weighted spherical harmonics to obtain
the modes $\psi_{\lm}$
and then solving 
$\ddot{h}_{\lm} = \psi_{\lm}$ using fixed frequency integration 
(FFI) \cite{Reisswig:2010di}. We represent this complex valued 
field in polar representation as
\begin{align}
  R h_{\lm} = A_{\lm} e^{- i \phi_{\lm}},
  \label{eq:polar_representation_gw_strain}
\end{align}
where $A_{\lm}, \phi_{\lm}$ are the amplitude and phase, respectively.
We plot all results against the retarded time coordinate
\begin{align}
  u = t - R_{\ast} =: t - R(r) - 2 M \log\left(\frac{R(r)}{2M}-1\right),
\end{align}
where $M$ is the total gravitational mass of the BNS system. 
$R(r) = r (1 + M/2r)^2$ is the radius in Schwarzschild coordinates
and $r$ corresponds to the radius in isotropic coordinates which 
we take to be the extraction radius of our simulation. The moment 
of merger $u_{\rm mrg}$ is estimated by looking at the dominant 
$(\ell,m)=(2,2)$ mode and the first peak of $A_{22}$ within the time 
frame where the merger is expected to occur.

The FFI applies a high-pass filter to remove non-linear drifts
generated by noise in the time integrations of $\ddot h_{\lm} = \psi_{\lm}$. 
Such a filter is characterized by a cut-off frequency
$\omega_{{\rm cut}\,\lm}$ for each mode.
We follow the suggestion made in \cite{Reisswig:2010di} to use 
$\omega_{{\rm cut}\,\lm} = m \omega_{0} / 2$, where $\omega_0=2\Omega_{0}$
is the GW frequency associated to the initial orbital angular frequency
$\Omega_{0}$.

An example of a waveform obtained from the ten-orbit inspiral 
\BAM{97} simulation using the \EFL{-WENOZ} scheme is presented 
in \autoref{fig:gw_bam97}. The wave 
train shows a first peak after around 21 cycles until merger 
near $u_{\rm mrg} = 2400 M$ where it is then followed up with 
a more complex structure that includes multiple peaks and a 
slow amplitude decay. We also plot the instantaneous GW frequency 
$\omega_{22} = - \mathcal{I}(\dot h_{22} / h_{22})$. It displays 
a drastic frequency increase near merger, which is a characteristic 
of a chirp-like signal.

We perform self-convergence studies based on simulations that 
use $(n_i) = (64, 96, 128, 160)$ 
points per direction on the highest refined AMR level, and name those resolutions as $(\textrm{LOW, MID, HIG, FIN})$.
We analyse the phase differences
\begin{align}
  \Delta \phi^{(n_i,n_j)}_{\lm} = \phi^{(n_i)}_{\lm} - \phi^{(n_j)}_{\lm}\,,
  \label{eq:phase_diff}
\end{align}
between pairs of resolutions. 
To determine the experimental convergence rate $p$
we rescale these differences by a factor $s$ that captures the rate by
which we expect the differences to decrease with increasing resolutions,
provided that our scheme converges. 
It is computed by \cite{Baumgarte:2010} 
\begin{align}
   s(p, n_i, n_j, n_k, n_l) &= \frac{1 - (n_i/n_j)^p}{(n_i/n_k)^p - (n_i/n_l)^p},
     \label{eq:convergence_scaling}
\end{align}
where $n_i < n_j \leq n_k < n_l$.

\begin{figure*}[t]
  \centering 
  \begin{tabular}[c]{cc}
  \includegraphics[width=0.49\textwidth]{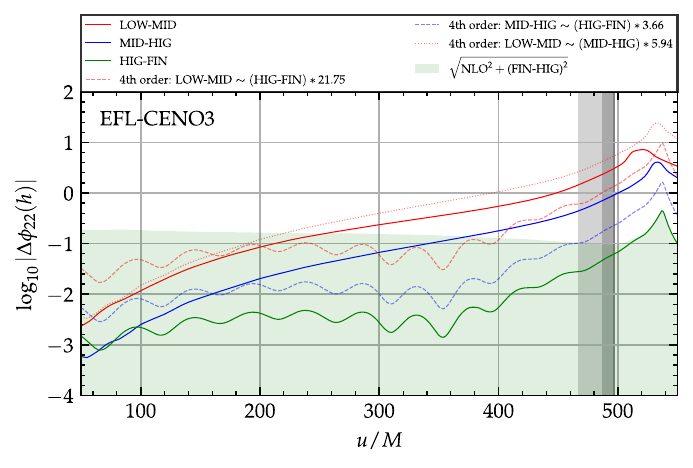}
  \includegraphics[width=0.49\textwidth]{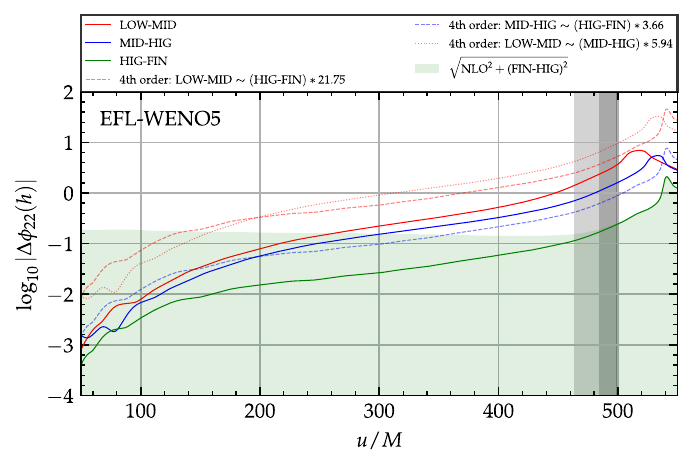}
  \end{tabular}
  \begin{tabular}[c]{cc}
  \includegraphics[width=0.49\textwidth]{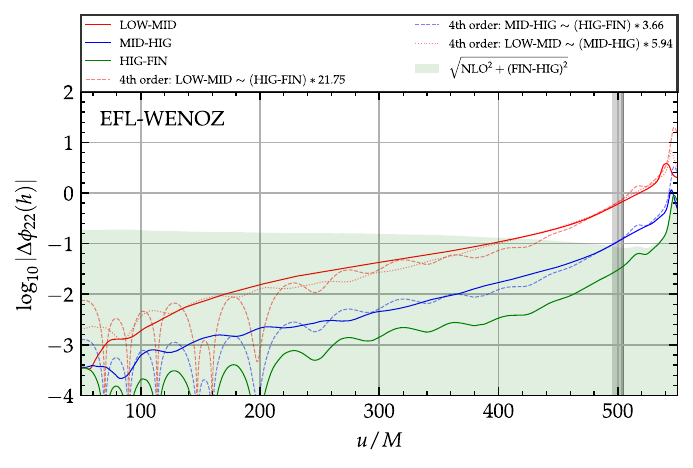}
  \includegraphics[width=0.49\textwidth]{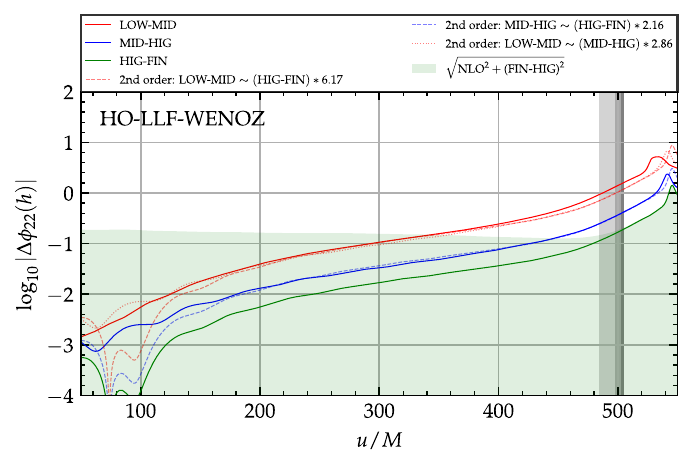}
  \end{tabular}
  \caption{GW phase difference convergence rate study for the 
  three-orbit \BAM{100} simulation: Solid lines represent phase 
  differences between runs with consecutively increasing resolutions; dashed 
  and dotted lines correspond to rescaled differences of HIG-FIN 
  and MIG-HIG differences, respectively, where the scaling factor 
  is computed from \eqref{eq:convergence_scaling} using an integer 
  convergence rate $p$; the latter was determined experimentally 
  by looking for the best visual overlap between dashed, dotted 
  and solid lines of the same color (which we failed to do with 
  the CENO3 and WENO5 reconstructions); the green shaded regions 
  indicate the obtained error budget as discussed in the text;
  the gray shaded regions mark the differences in merger times 
  between runs with consecutively increasing resolutions.}
  \label{fig:gw_phase_errors_bam100}
\end{figure*}

The error budget computation accounts for
(i) finite radius extraction errors $\Delta \phi^{\rm rad}_{\lm}$ 
and (ii) finite resolution errors $\Delta \phi^{\rm res}_{\lm}$.
Since they are of different origin and even come with a different 
sign \cite{Bernuzzi:2016pie} we compute the combined error using 
pointwise quadrature, 
\begin{align}
  \Delta \phi^{\rm err}_{\lm} = \sqrt{ 
    (\Delta \phi^{\rm rad}_{\lm})^2 + (\Delta \phi^{\rm res}_{\lm})^2 }.
  \label{eq:gw-error-budget}
\end{align}
The contribution (i) is estimated using the \textit{next-to-leading 
order} (NLO) behaviour of $\Psi_4$ \cite{Lousto:2010qx}. 
The contribution (ii) is estimated as the phase difference between the two highest
resolved runs, which we denote by (FIN-HIG).
The rational behind this is that, for a convergent scheme, any result 
obtained with higher precision as those runs will give results below 
this difference. 

\subsubsection{Three-orbit \BAM{100} simulation}

\autoref{fig:gw_phase_errors_bam100} shows a self-convergence study 
of the waveform phase differences for the \BAM{100} simulation. The 
first three panels correspond to the results obtained with the \EFL{} 
and the reconstruction schemes CENO3, WENO5 and WENOZ, respectively. 
The fourth panel shows results obtained with the hybrid algorithm 
HO-LLF-WENOZ. This last result serves as a reference for a comparison 
with our \EFL{} method.

\begin{figure*}[t]
  \begin{tabular}[c]{ccc}
  \includegraphics[width=0.49\textwidth]{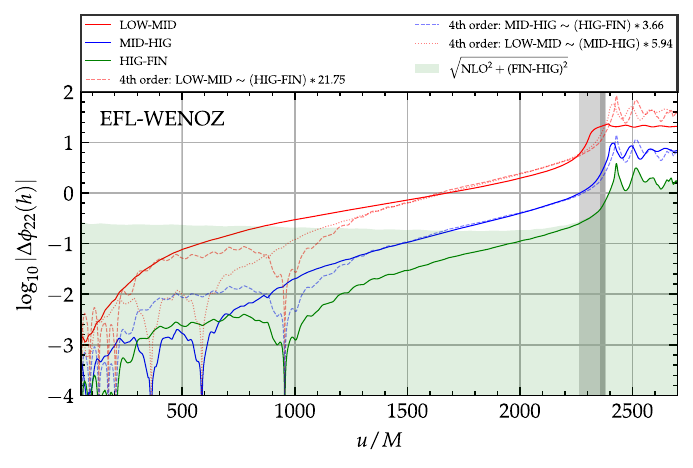}
  \includegraphics[width=0.49\textwidth]{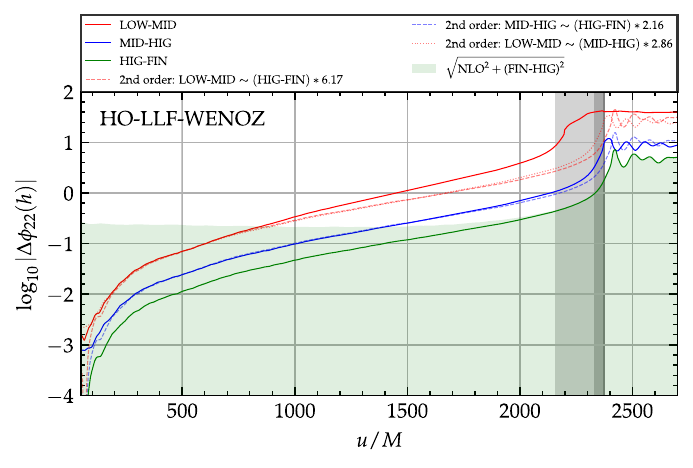}
  \end{tabular}
  \caption{GW phase difference convergence rate study for ten-orbit 
  \BAM{97} simulation. Left panel: \EFL{-WENOZ} method. Right panel: 
  HO-LLF-WENOZ method.}
  \label{fig:gw_phase_errors_bam97}
\end{figure*}

\begin{figure*}[t]
  \centering
  \begin{tabular}[c]{ccc}
    \includegraphics[width=0.49\textwidth]{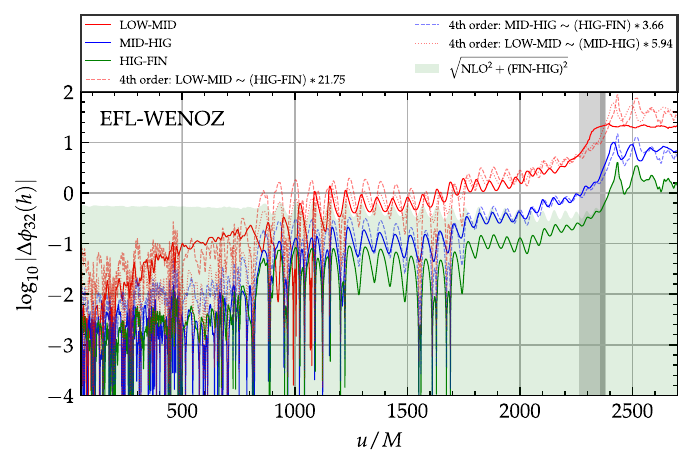}
    \includegraphics[width=0.49\textwidth]{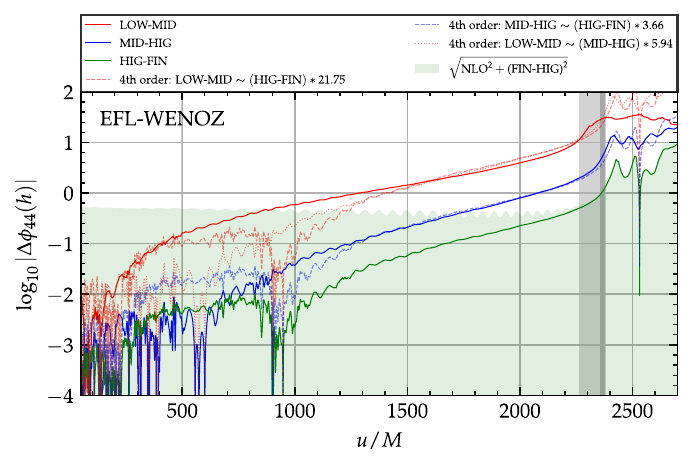}
  \end{tabular}
  \caption{GW phase difference convergence rate study for higher 
  modes for ten-orbit \BAM{97} simulation using the entropy flux 
  limiter with WENOZ reconstruction. Left panel: $(3,2)$ mode. 
  Right panel: $(4,4)$ mode.
  \label{fig:gw_phase_errors_bam97_higher_modes}}
\end{figure*}

The first thing to note is that the phase differences between runs 
with consecutively increasing resolutions decrease for all simulations.  
This indicates that the scheme is capable of providing consistent 
results. This is further supported by the decrease in the difference 
between merger times, indicated by the narrowing of the gray shaded 
regions, which marks the difference between merger times of runs with 
consecutively increasing resolutions. Furthermore, all runs converge 
to a merger time near $u = 500 M$ independent of the method. 
The figure contains the combined phase errors
\eqref{eq:gw-error-budget} as green shaded areas.
The estimated error of the highest resolved runs is ${\lesssim}0.1$~rad
uniformly throughout the inspiral phase and to merger. 

Focusing on the \EFL{} results for CENO3 and WENO5, one can see that, 
although the phase differences decrease, it is not possible to assign 
a clear integer convergence rate $p$ for which the rescaled differences 
match with the true differences (at the simulated resolutions).
By contrast, the plot demonstrates a
clear fourth order ($p=4$) convergence with the WENOZ method (bottom
left panel).
For the CENO3 series (top left panel) we actually find a scaling
consistent with $p = 2$ for the differences LOW-MID and MID-HIG, but a
higher scaling $p>2$ between the HIG-FIN difference.
In the WENO5 series (top right panel) the convergence plot is
strongly affected by the lowest resolutions, while the scaling seems
to be closer to $p=4$ between the MID-HIG and HIGH-FIN resolutions.
Also, the differences between resolutions for the WENO5 series are larger in absolute
values than those of other methods.
Overall, the difference between the CENO3, WENO5 and WENOZ series points to the
importance of the choice of reconstruction in the LO flux in Eq.~\eqref{eq:num_flx_split}.
In particular, the less dissipative and higher-resolution WENOZ scheme
(vs. WENO5) \cite{Borges:2008a} is a confirmed key feature in BNS
applications \cite{Bernuzzi:2012ci}.

Comparing the \EFL{} to the results obtained with the HO-LLF-WENOZ hybrid
we find that the former shows a faster convergence rate for the \EFL{-WENOZ}
series and generically smaller absolute differences HIG-FIN at merger
(except \EFL{-WENO5}, see above). The HO-LLF-WENOZ algorithm (bottom
right panel) yields a clean convergence pattern with $p = 2$, consistent 
with previous results reported in \cite{Bernuzzi:2016pie}, and phase
differences FIN-HIG (errorbars) at merger a factor ${\sim}5$ larger than \EFL{-WENOZ}.

\subsubsection{Ten-orbit \BAM{97} simulation}

\autoref{fig:gw_phase_errors_bam97} shows a convergence 
study similar to \autoref{fig:gw_phase_errors_bam100} but based on 
results of the ten-orbit \BAM{97} simulation. The simulations are
performed with \EFL{-WENOZ} and compared to those presented in
\cite{Bernuzzi:2016pie} and obtained with the HO-LLF-WENOZ method.

For both methods phase differences consistently decrease by increasing
the grid resolution. Both merger times tend to $u\approx2400 M$,
thus indicating the results are consistent (cf. \autoref{fig:proper_d_bam97}).
Convergence can be demonstrated clearly in both cases. The \EFL{-WENOZ}
scheme produces a clear fourth order ($p=4$) convergent waveforms,
consistent with the three-orbits simulations.
Instead, the HO-LLF-WENOZ scheme produces second order convergent
($p=2$) results starting at MID resolutions; the convergence
degrades for the LOW-MID difference towards the merger time \cite{Bernuzzi:2016pie}.
The phase differences FIN-HIG (errorbars) at merger for the
HO-LLF-WENOZ are a factor ${\sim}3$ larger than for the \EFL{-WENOZ}. In
both cases they are a factor ${\sim}10$ larger than in the three-orbit runs (for
comparable resolutions).
We also note that the convergence rate is maintained in the early
postmerger phase, suggesting that the \EFL{} scheme is robust and can
well capture the violent dynamics of the remnant NS. 

Given this clear convergence pattern for the $(2,2)$ modes of the \EFL{-WENOZ} runs,
we also investigate the convergence of higher modes $h_{\lm}$ with $\ell>2$. 
\autoref{fig:gw_phase_errors_bam97_higher_modes} shows a 
convergence study of the $(\l,m) = (3,2), (4,4)$ 
modes. Also in this case, the phase differences show a consistent decrease with
increasing resolution and a clear fourth-order convergence
of the modes' phase.
The phase error is of order $10^{-1}$~rad, with a flat
profile and rapidly accumulating only very close to the merger time.
Furthermore, the convergence pattern continues to hold through merger.
To our knowledge, this is the first time that a successful convergence study 
in higher modes of the GW strain is presented in the literature.

\section{Waveforms' Faithfulness}
\label{app:faithfulness}

\begin{table*}[t]
 \centering    
 \caption{Faithfulness functional $\mathcal{F}$ for the 
 BNS simulation considered in the present work, see 
 \autoref{tab:BNS_sim}. The values of $\mathcal{F}$ are 
 compared to the threshold values $\mathcal{F}_{\rm thr}$ 
 calculated with \eqref{eq:faithfulness} for different 
 signal-to-noise ratios. The notation 
 $\mathcal{F}_{\rm thr}^{\tilde \rho,\epsilon}=1-\epsilon^2/2\,\tilde{\rho}^2$ 
 has been introduced for the various threshold values, 
 which are explicitly presented on the second row inside 
 the brackets. A tick \cmark\, indicates that 
 $\mathcal{F} > \mathcal{F}_{\rm thr}$ and a cross \xmark\, 
 that $\mathcal{F} < \mathcal{F}_{\rm thr}$.}
   \begin{tabular}{cccccccccc}        
    \hline
    \hline
    Simulation & Scheme & $n$ & $\mathcal{F}$ & $\mathcal{F}_{\rm thr}^{14,6}$ & 
    $\mathcal{F}_{\rm thr}^{14,1}$ & $\mathcal{F}_{\rm thr}^{30,6}$ & 
    $\mathcal{F}_{\rm thr}^{30,1}$ & $\mathcal{F}_{\rm thr}^{80,6}$ & 
    $\mathcal{F}_{\rm thr}^{80,1}$ \\
    &&&& (0.9847) & (0.9974) & (0.9967) & (0.9994) & (0.9995) & (0.9999)\\
    \hline
    \BAM{97} & \EFL{-WENOZ}    & [160, 128] & 0.9998 & \cmark & \cmark & \cmark & \cmark & \cmark & \xmark \\
    \BAM{97} & HO-LLF-WENOZ & [160, 128] & 0.9992 & \cmark & \cmark & \cmark & \xmark & \xmark & \xmark \\
    \BAM{100}  & \EFL{-CENO3}    & [160, 128] & 0.9952 & \cmark & \xmark & \xmark & \xmark & \xmark & \xmark \\
    \BAM{100}  & \EFL{-WENO5}    & [160, 128] & 0.9991 & \cmark & \cmark & \cmark & \xmark & \xmark & \xmark  \\
    \BAM{100}  & \EFL{-WENOZ}    & [160, 128] & 0.9987 & \cmark & \cmark & \cmark & \xmark & \xmark & \xmark  \\
    \BAM{100}  & HO-LLF-WENOZ & [160, 128] & 0.9987 & \cmark & \cmark & \cmark & \xmark & \xmark & \xmark  \\
    \hline
    \hline
   \end{tabular}
 \label{tab:faithfulness}
\end{table*} 

We check if the \EFL{} numerical simulations are sufficiently 
accurate to produce faithful waveforms for gravitational-wave 
astronomy. We follow closely the methods and equations discussed 
in \cite{Damour:2010zb,Gamba:2020wgg}, to which we refer for 
a complete description. Previous results of this kind were 
presented in \cite{Bernuzzi:2011aq,Gamba:2020wgg}.

The accuracy of numerical waveform for application to GW 
astronomy is often quantified in terms of the {\it faithfulness} 
functional $\mathcal{F}\in [0,1]$ by considering criteria 
in the form \cite{Gamba:2020wgg,Damour:2010zb}:
\begin{equation}
 \label{eq:faithfulness}
  \mathcal{F} > \mathcal{F}_{\rm thr} = 1 - \frac{\epsilon^2}{2\,\tilde\rho^2},
\end{equation}
with $\epsilon^2 \leq 1$ and $\tilde\rho$ the signal-to-noise 
ration (SNR). Sometimes it is suggested \cite{Chatziioannou:2017tdw} 
to relax this criterion by taking $\epsilon^2=N$, where 
$N$ is the number of intrinsic parameters of the binary.
The criterion $\mathcal{F} > \mathcal{F}_{\rm thr}$ is a 
necessary condition that has to be satisfied by faithful 
waveform models, i.e.\ suitable for GW parameter estimation.
A possible violation of this criterion does not imply the 
presence of biases though. We compute threshold values 
$\mathcal{F}_{\rm thr}$ at SNRs $\tilde\rho=14$, $30$ 
and $80$ that correspond to the SNRs of GW190425, GW170817 
and a generic loud signal, respectively. For each of these 
SNRs the values of $\mathcal{F}_{\rm thr}$ are evaluated 
for two different choices of $\epsilon^2$, i.e.\ $\epsilon^2=1$ 
and $\epsilon^2=N=6$. The faithfulness $\mathcal{F}$ is 
evaluated using the numerical waveforms at two different 
resolutions. The faithfulness integral is computed over 
a frequency range $f \in [f_{\rm low},f_{\rm mrg}]$, where 
$f_{\rm low}$ corresponds to the initial circular GW frequency 
of the simulation\footnote{Note this corresponds to the 
first peak of the amplitude of  the Fourier transform of 
$\Re(h_{22})$.} and $f_{\rm mrg}$ is the merger frequency. 
We employ the \texttt{aLIGODesignSensitivityP1200087} 
\cite{TheLIGOScientific:2014jea} PSD from \texttt{pycbc} 
\cite{Nitz:2020} to compute the matches. In order to 
obtain accurate mismatch results from numerical data, 
one has to pre-process the raw $\psi_{4,\rm lm}$ modes 
before performing the FFI method to obtain $h_{\rm lm}$.
To this end we tapered the signals at the beginning and 
the end and also zero padded them for finer frequency 
bin resolution. The preprocessing should be done such 
that the instantaneous GW frequency $\omega_{22}$ computed 
from $h_{22}$ matches the GW frequency provided by the 
initial data, cf. table \ref{tab:BNS_sim}. We emphasize 
that this preprocessing step has no influence on the 
phase difference convergence rate.

\begin{figure}[t]
 \centering 
  \includegraphics[width=0.49\textwidth]{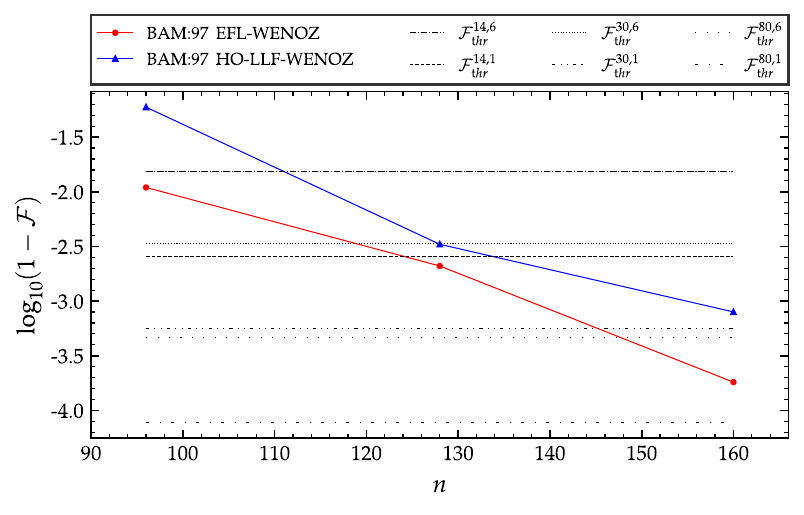}
  \caption{Faithfulness as a function of the resolution for 
  the \BAM{97} simulation.}
   \label{fig:faithfulness}
\end{figure}

\autoref{tab:faithfulness} reports the faithfulness 
values for the $(2,2)$ waveforms of the \BAM{97} and 
\BAM{100} simulations. Each value of $\mathcal{F}$ is 
obtained from the two highest resolution simulations 
available, that represent a measure of the error as 
discussed above. All the waveforms, except for \EFL{-CENO3}, 
produced with the \EFL{} method pass the three lowest 
accuracy criterion of \eqref{eq:faithfulness}. The 
same holds for the corresponding waveforms computed 
with the HO-LLF-WENOZ method of \cite{Bernuzzi:2016pie}.
Out of the six simulations examined only the \EFL{-WENOZ} 
for \BAM{97} passes a higher accuracy test than the 
one with SNR $\tilde\rho=30$ and $\epsilon^2=6$. Actually, 
this specific simulation at resolution of $n=160$ passes five out of the six  
accuracy tests making it an ideal candidate for GW modeling studies.
Note also that the faithfulness 
of \BAM{100} with \EFL{-WENO5} is very close to pass 
the fourth accuracy test $\mathcal{F}_{\rm thr}^{30,1}=0.9994$.

In \autoref{fig:faithfulness} we study the dependence 
of the faithfulness functional with simulation pairs of 
increasing resolution. Specifically, \autoref{fig:faithfulness} 
shows the faithfulness between pairs of waveforms at 
different resolutions $n=[160,128],[128,96]$ and $[96,64]$ 
as a function of the resolution. The plot focuses on 
the longest \BAM{97} simulation that is the most relevant 
for waveform modelling. It is apparent that for both 
schemes the quantity $1-\mathcal{F}$ converges, as 
expected, to zero with increasing resolution. Notice 
though that the EFL{-WENOZ} scheme produces more faithful 
waveforms than the HO-LLF-WENOZ scheme at the same 
resolution. With this convergence behaviour the EFL{-WENOZ} 
(HO-LLF-WENOZ) simulation is expected to pass the highest 
accuracy test $\mathcal{F}_{\rm thr}^{80,1}$ at resolution 
$n\sim 192$ ($n\sim 224$). The computational 
cost for a simulation at this resolution is approximately 
$\sim$1M CPU-hrs ($\sim$2.5M CPU-hrs).

\section{Conclusions}
\label{sec:conclusions}

This paper investigates, for the first time, the use of the \EFL{} scheme, 
an entropy-based flux-limiter for the computation of the hydrodynamics' numerical fluxes, 
in BNS merger simulations. 
The main question addressed here is whether the \EFL{} is sufficiently robust for the 
treatment of the NS surface and smooth-flow regions to provide us with accurate and 
high-order converging gravitational waveforms. We answer this in the affirmative. 

Our method builds on the proposal of \cite{Guercilena:2016fdl}, but notably 
does not make use of a positivity preserving limiter nor of free-parameters 
(see \autoref{sec:EFL_method}). The new \EFL{} scheme successfully passes 
a standard set of benchmark problems in special relativistic hydrodynamics, 
with results comparable to standard high-order characteristic WENO schemes, e.g.~\cite{Bernuzzi:2016pie}. Our scheme does not suffer from the oscillatory behaviour at the shock-tube 
discontinuities observed in the original implementation of \cite{Guercilena:2016fdl}.

Next, our method is tested in \autoref{sec:SNS} against 
three-dimensional general-relativistic single NS configurations. 
The \EFL{} scheme accurately locates the surface 
of stationary star solutions, see \autoref{fig:ideal_dyn_hybrid_plots}, 
and enables the use of the LO flux in this region while the interior
remains mainly resolved by the HO flux.
\EFL{} simulations give results comparable to those obtained with standard WENO schemes 
\cite{Radice:2013hxh,Bernuzzi:2016pie} and with the 
ELH \cite{Guercilena:2016fdl}. However, our new 
simulations are free of the spurious direction-dependent effects found in
\cite{Guercilena:2016fdl}, see \autoref{fig:ideal_rho_max_dyn}.
In addition, the \EFL{} scheme performs very well in the simulation
of rapidly rotating stars. As shown by
\autoref{fig:vy_profile_dyn_ideal_rot}, the velocity profile and the
sharp transition at the surface of the star are almost un-altered over 
four rotational periods and correctly converge to the exact (initial)
solution. The results from both high-order WENO scheme or second-order 
finite-volume schemes with primitive reconstruction are significantly 
less accurate (at the same resolutions).

Finally, in \autoref{sec:BNS}, the new entropy method is 
applied to BNS merger simulations. The EFL scheme can be used to successfully 
evolve binaries and the properties observed in single star tests carry over 
to the simulation with non-stationary spacetimes and neutron stars moving on 
the computational grid. As shown in \autoref{fig:hybrid_plots_gamma} 
and \autoref{fig:hybrid_plots_sly}, the entropy limiter locates the surface of the 
inspiraling NSs quite accurately and it converges to zero in regions of smooth flows. 
Further, it captures the collisional shocks at merger and the outward dynamics of spiral 
density waves, thus being robust also for postmerger evolutions.

A convergence study of the gravitational waveforms obtained from these simulations 
shows that the \EFL{} with a low-order flux based on the WENOZ reconstruction (\EFL{-WENOZ}) 
can deliver fourth-order convergent waveforms at current production resolutions 
(\autoref{sec:gw_analysis}). Such a convergence is measured in the $(2,2)$ dominant 
mode of the strain but also in the next subdominant modes $(3,2)$ and $(4,4)$. 
To our knowledge, these are the first results in which fourth-order convergence 
is demonstrated. The estimated phase error in the \EFL{-WENOZ} waveform is about a 
factor ${\sim}5$ smaller than the error in the state-of-the-art high-order WENOZ scheme used in 
the same BAM code, at the same resolution.

We conclude that our \EFL{} scheme can be efficiently used for high-quality waveform 
production and for future large-scale investigations of the binary NS parameter space. 
These studies will aim at extending our previous investigation in both quality and 
simulation length 
\cite{Bernuzzi:2011aq,Bernuzzi:2012ci,Bernuzzi:2013rza,Dietrich:2015iva,Dietrich:2015pxa}. 
The immediate target is to resolve tidal effects near the merger that are the main source of 
systematic error in current waveform approximations of GW astronomy \cite{Gamba:2020wgg}. 
We estimate that this will require \EFL{-WENOZ} multi-orbit and multi-resolution simulations 
resolving the NSs up to $n\sim 192$ grid points per direction. A ten-orbit convergent series is 
within reach of modern supercomputers (similar to those used for this work) at the 
approximate cost of ${\sim}$2M CPU-hrs.

In the postmerger regime, the EFL well tracks the front of the 
ejecta. From the right panels of \autoref{fig:hybrid_plots_gamma} 
and \autoref{fig:hybrid_plots_sly} it is apparent that the EFL 
is triggered by the outward dynamics of the spiral density waves. 
The main source of inaccuracy of the ejecta is the progressively 
lower resolution of the low density material as it propagate 
outwards. There might be a benefit in using the high-order 
scheme of the EFL when ejecta starts to propagate but, most 
likely, the re-definement of the grid will become too severe 
at very large distances and it is likely the EFL performs similarly 
to other schemes. A detailed investigation of the benefits of 
the EFL for resolving the ejecta is left to further investigation.

\begin{acknowledgements}
  We would like to thank members of the Jena group for fruitful 
  discussions and invaluable input. Especially, we would like 
  to thank Francesco Maria Fabbri for help with the TOV and RNS 
  simulations and Rossella Gamba for kindly providing her scripts 
  for the computation of the faithfulness functionals. We thank 
  Federico Guercilena and David Radice for comments on the manuscript.
   
  G.~D. and S.~B. acknowledge support by the EU H2020 under ERC 
  Starting Grant, no.~BinGraSp-714626. G.~D. is co-financed by 
  Greece and the European Union (European Social Fund-ESF) through 
  the operational programme ``Human Resources Development, Education 
  and Lifelong Learning'' in the context of the project ``Reinforcement 
  of Postdoctoral Researchers-2nd Cycle'' (MIS-5033021), implemented 
  by the State Scholarships Foundation (\href{http://www.iky.gr}{IKY}). 
  F.~A. was supported in part by the Deutsche Forschungsgemeinschaft 
  (DFG) under Grant No. 406116891 within the Research Training Group
  RTG 2522/1. 
  
  Computations where performed on the national HPE Apollo Hawk 
  at the High Performance Computing Center Stuttgart (HLRS), 
  on the ARA cluster at Friedrich Schiller University Jena 
  and on the supercomputer SuperMUC-NG at the Leibniz-Rechenzentrum 
  (LRZ, \url{www.lrz.de}) Munich. The ARA cluster is funded 
  in part by DFG grants INST 275/334-1 FUGG and INST 275/363-1 
  FUGG, and ERC Starting Grant, grant agreement no.~BinGraSp-714626. 
  The authors acknowledge HLRS for funding this project by 
  providing access to the supercomputer HPE Apollo Hawk under 
  the grant number INTRHYGUE/44215. The authors acknowledge 
  also the Gauss Centre for Supercomputing e.V. 
  (\url{www.gauss-centre.eu}) for funding this project by 
  providing computing time to the GCS Supercomputer SuperMUC-NG 
  at LRZ (allocations pn56zo, pn68wi).

\end{acknowledgements}

%%______________________________________________________________


\end{document}